\documentclass{emulateapj}

\newcommand{\figref}[1]{Figure \ref{#1}}
\newcommand{\eqref}[1]{equation (\ref{#1})}

\newcommand{\sub}[1]{\ensuremath{_{\mbox{\scriptsize#1}}}}

\begin{document}
\slugcomment{To appear in ApJ}
\title{Planet Shadows in Protoplanetary Disks. I: Temperature Perturbations}
\author{Hannah Jang-Condell\altaffilmark{1}}
\affil{Carnegie Institution of Washington, Department of Terrestrial Magnetism,
Washington, D.C. 20015}
\email{hannah@astro.umd.edu}
\altaffiltext{1}{Current address: 
Department of Astronomy, University of Maryland, College Park,
MD, 20740-2421}
\shorttitle{PLANET SHADOWS I. TEMPERATURE PERTURBATIONS}
\shortauthors{H. JANG-CONDELL}

\begin{abstract}
Planets embedded in optically thick passive accretion disks 
are expected to produce perturbations in the density and temperature 
structure of the disk.  We calculate the magnitudes of these 
perturbations for a range of planet masses and distances.  
The model predicts the formation of a 
shadow at the position of the planet paired with a brightening 
just beyond the shadow.  
We improve on previous work on the subject by 
self-consistently calculating the temperature and density structures 
under the assumption of hydrostatic equilibrium and taking the full 
three-dimensional shape of the disk into account rather than assuming 
a plane-parallel disk.  
While the excursion in temperatures is less 
than in previous models, the spatial size of the perturbation is larger.  
We demonstrate that a 
self-consistent calculation of the density and temperature structure 
of the disk has a large effect on the disk model. In addition,  
the temperature structure in the disk is 
highly sensitive to the angle of incidence of stellar irradition at the 
surface, so accurately calculating the shape of the disk surface is 
crucial for modeling the thermal structure of the disk.  
\end{abstract}

\keywords{planetary systems: formation --- 
planetary systems: protoplanetary disks ---
radiative transfer}

\section{Introduction}

Giant planets forming by core accretion need to have cores of 
10-20 M$_{\earth}$ to be massive enough to 
accrete a gaseous envelope \citep{2005Hubickyj_etal}.  
This predicts that sizeable planet embryos form before 
circumstellar gas disks dissipate.  
These disks are typically modeled as passive accretion disks, 
optically thick and gas-dominated.  Their temperature structure 
is strongly dependent on heating from stellar irradiation
\citep{CG, calvet, vertstruct, dalessio2, dalessio3}.  
In particular, the disk temperatures depend strongly 
on the angle of incidence of the stellar irradiation on the 
disk surface.  

While a substantial amount of work on numerical hydrodynamic 
simulations of planets embedded in disks has been carried out, 
calculating the radiative transfer of stellar irradiation 
in conjunction with these simulations is 
prohibitively difficult.  
Simulations by \citet{bate} illustrate the effect of 
hydrodynamics on the disk structure around a small embedded planet, 
but use an isothermal equation of state.  
The effects of MHD turbulence have been studied by 
\citet{2004MNRAS.350..829P} and \citet{2007Oishi_etal}, 
but again, assuming a simple and unrealistic equation of state.  
\citet{2006KlahrKley} and 
\citet{2006PaardekooperMellema,2008PaardekooperMellema} 
have made an effort to include radiative transfer as well as hydrodynamics 
in their calculations, but do not include the effects of stellar irradiaton 
on their models.  

The work presented in this paper does not include hydrodynamics, 
but rather focuses on the effects of stellar irradiation, 
which is a particularly important source of heating 
in the photospheres of disks.  Since it is the photospheres of 
optically thick disks that are observed, the effects of stellar irradiation 
at the surface of disks are an important consideration in predicting 
observations of protoplanetary disks.  A limitation of the hydrodynamic 
simulations described above is that in order to adequately model 
the densest regions of the gas they are necessarily limited in 
spatial range above the midplane.  The photosphere and surface of 
a disk are often several scale heights above the midplane.  The models 
presented here are complementary to detailed hydrodynamic simulations 
for this reason.  

We are particularly interested 
in determining if the growing cores of giant planets produce effects 
that are observable.  If these effects are observed, they would 
affirm the core accretion model as the paradigm for giant planet formation.  
While a fully-formed Jupiter-mass planet would produce a larger 
feature in a disk, it would reveal little about how it formed.  

Previous work on small planets embedded in optically thick 
gas disks indicates that sub-Jovian mass planet cores 
can perturb the disk enough to alter the temperature structure 
of the disk in the immediate vicinity of the planet, 
with consequences for further evolution of the planet
\citep{paper1, paper2, HJCmigration}.  
In \citet{paper1,paper2} (henceforth Paper I and Paper II, respectively), 
the planet is predicted to gravitationally compress the disk in the vertical 
direction, creating a shadow paired with a bright spot, leading 
to temperature variations.  However, those models 
were limited by being plane-parallel, despite the sizes of the 
simulation boxes used.  In addition, while the density perturbations 
were calculated under hydrostatic equilibrium, they were not 
calculated self-consistently with the temperature perturbations.  

In this paper, we improve upon the model presented in Papers I and 
II by iteratively calculating the density and temperature structure 
of the disk for self-consistency and eliminating the assumption of 
a locally plane parallel model.  
The paper is organized as follows:
in \S \ref{model} we describe the basic model and the improvements 
we have made,
in \S \ref{stellarheating} we elaborate in detail on the method we 
use for modeling radiative transfer, 
in \S \ref{results} we describe our results for a different planet 
masses and distances, 
in \S \ref{discussion} we discuss our results in comparison to 
previous models, and 
in \S \ref{conclusion} we present our conclusions.

\section{Model Calculation}\label{model}

We adopt the formalism developed in \citet{paper1,paper2}
(henceforth, Paper I \& II, respectively) 
for calculating the effects on temperature of shadowing and brightening 
of stellar illumination at the disk's surface.  
These methods are based on those of \citet{calvet} and 
\citet{vertstruct,dalessio2} in a 1-D plane-parallel disk.  

The improvements made on the previous model include 
iteratively calculating the temperature and density structure of the disk 
for self-consistency.  
The new model is not plane-parallel as the previous models were.  
It accounts for the variation of disk structure with radius and 
the curvature of the disk in azimuth.  

Each iteration proceeds as follows.  
Starting with the density structure of the disk, 
we calculate the temperature of the disk in radiative equilibrium 
with viscous heating and stellar irradiation as heating sources.  
We adapt methods developed in Papers I and II to calculate 
radiative transfer on the surface of a perturbed disk.  Then, 
given the new temperature structure, 
we recalculate the density structure assuming vertical hydrostatic
equlibrium.  

Each of these steps are  described in detail below.  
The initial conditions, which assume azimuthal symmetry, 
are calculated iteratively until they 
reach stability.  The computationally intensive nature of the calculations 
prohibits large numbers of iterations when a planet is added, so 
relatively few iterations are carried out for the planet in the disk.  

\subsection{Model Parameters}

We use rotating cyclindrical coordinates $(r,\phi,z)$ throughout, 
with the star at the origin, the planet at $(a, 0, 0)$, and 
the $z$-axis aligned with the orbital angular momentum vector.  
We adopt parameters 
for the stellar mass, radius, and effective temperature of 
$M_*=1\,M_{\sun}$, $R_*=2.6\,R_{\sun}$, and $T_*=4280$ K, 
corresponding to an age of 1 Myr \citep{siess_etal}.  
The mass accretion rate is $\dot{M} = 10^{-8}\,M_{\sun}\mbox{yr}^{-1}$, 
and the viscosity parameter is $\alpha_v=0.01$, which are 
typical of T-Tauri type stars.

The simulation box is centered on the planet at $(a,0,0)$.  
The size of the box is scaled relative 
to the Hill radius, 
$r\sub{Hill} = a(m_p/3M_*)^{1/3}$.  
The box spans from $r\sub{min}=a-12.5r\sub{Hill}$ to 
$r\sub{max}=a+12.5r\sub{Hill}$ in $r$, 
and from $\phi\sub{min}=-12.5r\sub{Hill}/a$ to 
$\phi\sub{max}=12.5r\sub{Hill}/a$ in $\phi$.  
The height of the box is set to $z\sub{max}\sim 2z_s$ where $z_s$ is the 
height of the unperturbed disk surface at $r=a$.  
We assume symmtery across the midplane, so $z\sub{min}=0$.  
The box is decomposed onto a grid of $100\times100\times100$ points  
equally spaced in $r$, $\phi$, and $z$, with 
the grid slice at $\phi=\phi\sub{min}$ used as a placeholder 
for structure of the unperturbed disk.   

We use the opacities from \citet{dalessio3} using a dust model with 
parameters $a\sub{max} = 1\,\mbox{mm}$, $T=300$ K, and $p = 3.5$, 
assuming that the dust opacities are constant throughout 
the disk.  The values for the opacities (in $\mbox{cm}^2 \mbox{g}^{-1}$)
are as follows: the Rosseland mean opacity is $\chi_R=1.91$, 
the Planck mean opacity integrated over the disk spectrum (300 K) is $\kappa_P=0.992$,
and the Planck mean opacities integrated over the stellar spectrum (4000 K) are 
$\kappa_P^*=1.31$ for absorption alone and $\chi_P^*=5.86$ for absorption plus scattering.  
The absorption fraction is then 
$\alpha\sub{abs} = \kappa_P^*/\chi_P^*$, 
while the scattered fraction is $\sigma = 1-\alpha\sub{abs}$.  
The Rosseland mean opacity is used to calculate the photosphere of the 
disk, and $\chi_P^*$ is used to calculate the surface of the disk.  

\subsection{Heating Sources}

The basic calculation for radiative heating is described in
detail in \citet{paper1}.  Here, we summarize those methods.  
The two main heating sources are 
from viscous heating ($\Gamma_v$) 
and stellar irradiation ($\Gamma_r$), 
offset by radiative cooling,
\begin{equation}
\Lambda = \rho\chi_R\sigma_B T^4, 
\end{equation}
where $\rho$ is density, 
$\sigma_B$ is the Stefan-Boltzmann constant and 
$T$ is the local temperature of the gas.  
Under equilibrium conditions, 
\begin{equation}\label{totalheat}
\Gamma_v + \Gamma_r = \Lambda.
\end{equation}
If $T_v$ and $T_r$ give the equilibrium temperature for solely 
viscous heating or stellar irradiation, respectively, 
then it follows that 
the equilibrium temperature given both sources of heating is 
\begin{equation}
T\sub{eq} = (T_v^4 + T_r^4)^{1/4}.
\end{equation}

\subsubsection{Viscous Heating}

We assume that viscous flux is generated at the midplane and 
transported radiatively in a grey atmosphere so that 
\begin{equation}
T_v = \left[\frac{3F_{v,\mbox{\scriptsize ph}}}{8\sigma_B}(\tau_d+2/3)\right]^{1/4}.
\end{equation}
where $\tau_d$ is the optical depth perpendicular to the disk 
using the Rosseland mean opacity, $\chi_R$: 
\(\tau_d = \int_z^{\infty} \chi_R \rho \, dz'\).
The viscous flux emitted at the photosphere
$F_{v,\mbox{\scriptsize ph}}$ at a distance $r$ for a star of 
mass $M_{\star}$ and radius $R_{\star}$ 
accreting at a rate $\dot{M}_a$ is 
\begin{equation}
F_{v,\mbox{\scriptsize ph}} = \frac{3GM_{\star}\dot{M}_a}{4\pi r^3}
	\left[1-\left(\frac{R_{\star}}{r}\right)^{1/2}\right]
\end{equation}
\citep{pringle}.  
Viscous heating above the surface is assumed to be negligible.  

\subsubsection{Stellar Irradiation}

The amount of heating from stellar irradiation is calculated 
using methods developed in Papers I and II for a plane-parallel 
disk and adapted for a fully three-dimensional system.  The full 
details of this calculated are rather lengthy, so this has been 
set out into \S\ref{stellarheating}.

\subsubsection{Differential Rotation}

The gas moves in streamlines past the planet 
at approximately the Keplerian rate, resulting in the shearing out of the 
hot and cold spots as material moves into and out of shadows and 
brightenings.  The calculation is done in steady state, meaning that 
the gas is treated as a steady flow so that 
temperatures as a function of spatial position is constant 
even though the gas itself is in motion.  
Given the bulk velocity along a 
particular streamline, the movement of the gas from one grid cell 
to the next is equivalent to a time step equal to the length of the 
grid cell divided by the velocity.  

The equilibrium temperature 
calculated in \eqref{totalheat} represents the total 
amount of heating due to disk viscosity and stellar irradiation.  
We assume that the specific heat per unit surface area of the disk is 
$k\Sigma/\bar{m}$, where $k$ is the Boltzmann constant 
and $\bar{m}$ is the mean molecular weight of the gas, 
which we assume to be primarily molecular hydrogen.  
Then we can approximate the rate at which 
a parcel of gas radiatively heats or cools as 
\begin{equation}
\frac{k\Sigma}{m} \frac{\partial T}{\partial t} = 
\sigma_B (T\sub{eq}^4 - T^4).
\end{equation}
If $T\sub{eq}<T$, the parcel of gas cools, 
if $T\sub{eq}>T$ the parcel of gas heats.  

\subsection{Density Profile}
To calculate the density profile for the disk, we assume 
hydrostatic equilibrium, 
\begin{equation}
\frac{1}{\rho}\frac{dP}{dz} = -g_z, \label{HSE}
\end{equation}
where $\rho$ is the density, $P$ is the pressure, 
and $g_z$ is the $z$-component of gravity.  
We assume the ideal gas law, 
$P=\rho kT/\bar{m}$.  
In the absence of the planet, the sole contribution to the gravity is the star:
\begin{equation}
g_z = \frac{G M_* z}{(r^2+z^2)^{3/2}}. \label{grav_unpert}
\end{equation}
We use this equation to calculate the initial conditions.  
When a planet is added, the gravity has an additional contribution 
from the planet, so \eqref{grav_unpert} becomes 
\begin{equation}
g_z = \frac{G M_* z}{(r^2+z^2)^{3/2}} 
+ \frac{G M_p z}{(r^2+r_p^2-2r\,r_p\cos\phi+z^2)^{3/2}}.
\end{equation}

Given a vertical temperature profile, we calculate the density profile by 
integrating \eqref{HSE} from the top of the simulation box down the midplane.  
For the perturbed disk, we require conservation of the total surface 
density.  
In a standard viscous accretion disk, the surface density of the disk 
and temperature of the disk are coupled.  Thus, for the initial disk 
we normalize the total integrated surface density 
\begin{equation}\label{surfden_int}
\Sigma = 2 \int_{0}^{\infty} \rho\, dz'
\end{equation}
with the surface density given by a steadily accreting viscous disk
\begin{equation}\label{surfden_vis}
\Sigma = \frac{\dot{M}}{3\pi\nu_v} 
	\left[1-\left(\frac{R_{\star}}{r}\right)^{1/2}\right] 
\end{equation}
where $\nu_v$ is the viscosity of the disk \citep{pringle}.
We adopt a standard Shakura-Sunyaev viscosity \citep{shaksun} with 
$\nu_v = \alpha_v c_0 H$
where $\alpha_v$ is a dimensionless parameter, $c_0$ is the 
sound speed at the midplane, and $H$ is the thermal scale height,  
given by $H=c_0/\Omega_K$ where $\Omega_K$ is the Keplerian angular velocity.

\section{3D Radiative Transfer of Stellar Irradiation}\label{stellarheating}

In this section, we describe the details of radiative transfer of 
stellar irradiation on the surface of a disk perturbed by a planet.  
These methods are based on the work 
of \citet{calvet} and \citet{vertstruct} which calculated the vertical 
temperature structure of a locally plane-parallel disk model without 
perturbations.  
In Papers I and II, we developed a method for using their solutions 
to calculate radiative transfer on a perturbed locally plane-parallel 
disk.  Here, we extend the formalism to a fully three-dimensional disk.  

For a plane-parallel medium, the optical depth at disk-temperature 
frequencies perpendicular to the surface ($\tau_d$)
is related to the line-of-sight optical depth to stellar radiation 
($\tau_s$) as \(\tau_s = \chi_P^*\tau_d/(\chi_R\mu)\) 
where $\mu$ is the angle of 
incidence of the stellar radiation at the surface.  
As shown in Paper II, 
$T_r$ for a locally plane-parallel disk is given by 
\begin{equation}\label{irradflux}
B(\tau_s) = 
\frac{\sigma_B T_r^4}{\pi} = \frac{\alpha\sub{abs} F\sub{irr} \mu}{4\pi}
   \left[ c_1 + c_2 e^{-\tau_s} + c_3 e^{-\beta\mu\tau_s} 
     \right], 
\end{equation}
where 
$\mu$ is the cosine of the 
angle of incidence of stellar irradiation at the surface 
of the disk, 
$\beta \equiv \sqrt{3\alpha\sub{abs}}$, 
and the stellar flux incident at the surface is 
\begin{equation}
F\sub{irr} = \frac{\sigma_B T_*^4 R_*^2}{(r_s^2+z_s^2)}
\end{equation}
where $(r_s,\phi_s,z_s)$ are the coordinates at the surface.
The remaining coefficients are 
\begin{eqnarray}
c_1 &=& \frac{ 6 + 9\mu\chi_R/\chi_P^* }{\beta^2}
- \frac{6 (1-\chi_R/\chi_P^*)\left(3-\beta^2\right)}{\beta^2
   (3 + 2 \beta) (1 + \beta \mu)} \label{c1}
\\
c_2 &=& 
\left(\frac{\chi_P^*}{\mu \kappa_P}-\frac{3\mu\chi_R}{\chi_P^*}\right)
\frac{(1-3\mu^2)}{(1-\beta^2 \mu^2)} \label{c2}
\\
c_3 &=& \left(\frac{\beta \chi_P^*}{\kappa_P}-\frac{3\chi_R}{\chi_P^*\beta}\right)
\frac{(2+3\mu)(3-\beta^2)}{\beta(3+2\beta)(1-\beta^2 \mu^2) }. \label{c3}
\end{eqnarray}

We adapt these equations to a three-dimensional disk by dividing 
the surface into grid elements by $r_i$ and $\phi_j$ and numerically 
integrating the contributions from each surface element. 

\subsection{Flux Contributions}

For a surface element $\delta A$ located at $S$, 
the contribution to the 
flux at point $P=(r,\phi,z)$ in the disk is proportional to 
$B(\tau_s^{PS})$ as given in equation \ref{irradflux}, 
with the optical depth calculated as 
\begin{equation}
\tau_s^{PS} = 2/3 + \frac{\nu}{\mu}\int_P^S \chi_P^*\rho\,dl
\end{equation}
where the integral is carried out along the line segment $\overline{PS}$ 
and $\nu$ is the cosine of the angle between $\overline{PS}$ and the 
surface element.  The $2/3$ term on the left hand side of the equation 
accounts for the optical depth from the star to the surface.  
The second term on the left hand side of the equation is what 
$\tau_s$ would be between $P$ and the infinite plane represented 
by the surface element.  
Defining $d\Omega$ to be 
the solid angle subtended by the surface element, 
then the total heating from stellar irradiation at $P$ 
in the limit as $d\Omega \rightarrow 0$ is 
\begin{equation}\label{B_integral} 
B\sub{tot}(P) = \frac{1}{\pi}\int B(\tau_s^{PS},\mu)\, \nu\, d\Omega.  
\end{equation}
For a plane parallel disk, this equation reverts to 
equation \ref{irradflux}.  
To calculate this numerically over an assemblage of finite surface 
elements, we assume that $B(\tau_s^{PS},\mu)$ stays fairly constant 
over a given surface element, and equation \ref{B_integral} becomes 
\begin{equation}\label{B_sum}
B\sub{tot}(P) \approx \frac{1}{\pi}\sum_{\forall S} B(\tau_s^{PS},\mu)\, 
\int_{S}\nu\, d\Omega.  
\end{equation}
The expression for $\int_{S} \nu\,d\Omega$ integrated over an arbitrary 
triangle with respect to $P$ can be determined analytically 
as follows.  We define 
$\mathbf{s}_1$,
$\mathbf{s}_2$, and 
$\mathbf{s}_3$, 
to be the vectors that point from $P$ to the vertices of the triangle.  
Let $\hat\mathbf{n}$ be the unit normal to the triangle and $\theta_{ij}$ 
be the angle begin the vectors $\mathbf{s}_i$ and $\mathbf{s}_j$.
Then
\begin{eqnarray}\label{momentsolidangle}
\int_{S} \nu\,d\Omega &=& \hat{\mathbf{n}} \cdot \left[
\frac{\theta_{12}}{\sin\theta_{12}} (\mathbf{s}_1 \times \mathbf{s}_2)
+\frac{\theta_{23}}{\sin\theta_{23}} (\mathbf{s}_2 \times \mathbf{s}_3)
\right.\nonumber \\
&&
\hspace{7em}\displaystyle\left.
+\frac{\theta_{31}}{\sin\theta_{31}} (\mathbf{s}_3 \times \mathbf{s}_1)
\right].
\end{eqnarray}

\begin{figure}[htbp]
\plotone{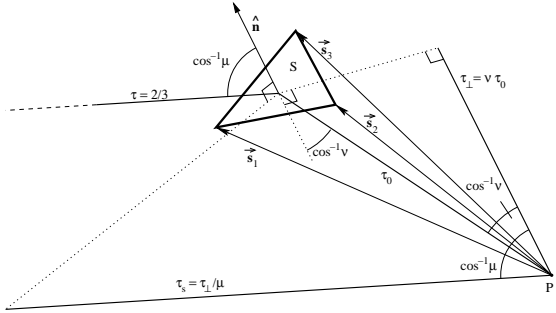}
\caption{\label{solidangle}Angles used in calculating optical depths}
\end{figure}

\subsection{Surface Decomposition}

We define the surface of the disk to be where 
the optical depth to stellar irradiation integrated along 
the line of sight is $\tau_s=2/3$.  
The optical depth at a given point $P=(r,\phi,z)$ in the disk is 
\begin{equation}
\tau_s(r,\phi,z) = \tau\sub{in} + \int_{\ell} \chi_P^*\rho\,dl,
\end{equation}
integrating along the line segment $\ell$, which extends from the point toward 
the star, ending at the inner boundary of the 
box, at $(r\sub{min},\phi,z\sub{in})$
where $z\sub{in}=z r\sub{min}/r$.  
The optical depth at the inner edge of the simulation box, $\tau\sub{in}$, is 
assumed to be 
\begin{equation}
\tau\sub{in} = \frac{1}{2}\chi_P^*\rho(r\sub{min},\phi,z\sub{in})
               \sqrt{r\sub{min}^2+z\sub{in}^2}.
\end{equation}
The density increases monotonically toward the midplane, so for 
every coordinate pair $(r_i,\phi_j)$ the 
height of the surface 
$h_{i,j}$ can be uniquely determined.  
The surface is defined by the set of points 
$\mathbf{s}_{i,j} = (r_i,\phi_j, h_{i,j})$.

The set of neighboring points, 
$\mathbf{s}_{i,j}$, $\mathbf{s}_{i+1,j}$, $\mathbf{s}_{i+1,j+1}$, and $\mathbf{s}_{i,j+1}$,  
define a surface element over which radiative transfer from 
stellar irradiation is calculated.  The midpoint of this set of 
four points is 
\begin{eqnarray*}
\mathbf{s}_m &=& \left[ 
\left(\frac{r_i+r_{i+1}}{2}\right)\cos\left(\frac{\phi_{j+1}-\phi_j}{2}\right),
\frac{\phi_j+\phi_{j+1}}{2}, \right. \\
&&\hspace{7em}\left.\frac{h_{i,j}+h_{i+1,j}+h_{i+1,j+1}+h_{i,j+1}}{4}
\right]
\end{eqnarray*}
For the unperturbed disk $h_{i,j}$ is independent of $j$, and 
$\mathbf{s}_{i,j}$, $\mathbf{s}_{i+1,j}$, $\mathbf{s}_{i+1,j+1}$, and $\mathbf{s}_{i,j+1}$ 
are co-planar.  If $\hat{\mathbf{n}}$ is the 
normal to the plane through these four points, then the cosine of the 
angle of incidence is 
\begin{equation}\label{mu}
\mu = \hat{\mathbf{s}}_m \cdot \hat{\mathbf{n}} + \frac{4 R_*}{3\pi|\mathbf{s}_m|}
\end{equation}
where the second term on the right side of the equation 
is the minimum allowed value of $\mu$, resulting 
from the finite size of the star.  Note that in Papers I and II, we neglected 
to include this term in calculating $\mu$, which resulted 
in more marked shadowing in the disk.  

For the perturbed disk, the points 
$\mathbf{s}_{i,j}$, $\mathbf{s}_{i+1,j}$, $\mathbf{s}_{i+1,j+1}$, and $\mathbf{s}_{i,j+1}$
are not necessarily co-planar.  In this case, we subdivide the surface element 
into four triangles defined by the midpoint 
and two adjacent points, 
i.e.~$(\mathbf{s}_{i,j}, \mathbf{s}_{i+1,j}, \mathbf{s}_m)$,
$(\mathbf{s}_{i+1,j}, \mathbf{s}_{i+1,j+1}, \mathbf{s}_m)$,
$(\mathbf{s}_{i+1,j+1}, \mathbf{s}_{i,j+1}, \mathbf{s}_m)$,
and $(\mathbf{s}_{i,j+1} \mathbf{s}_{i,j}, \mathbf{s}_m)$, 
with normal vectors 
$\hat{\mathbf{n}_1}$, $\hat{\mathbf{n}}_2$, $\hat{\mathbf{n}}_3$, and $\hat{\mathbf{n}}_4$, respectively.  The angles of incidence are calculated as 
\(\mu_i = \hat{\mathbf{s}}_m \cdot \hat{\mathbf{n}}_i + 4 R_*/(3\pi|\mathbf{s}_m|) \), 
and the contributions to radiative heating are summed over each triangle 
individually, with $F\sub{irr}$ calculated at $\mathbf{s}_m$ for all 
of them.

In order to avoid a discontinuity of temperature above the surface of 
the disk, the temperature above the surface is taken to be 
$B(\tau_s,\mu)$ where $\mu$ is now the angle of incidence to the surface of 
constant $\tau_s$:
\begin{equation}
\mu = \frac{\mathbf{r}\cdot\nabla\tau_s}{|\mathbf{r}||\nabla\tau_s|}.
\end{equation}

Surface elements outside the simulation box are assumed to follow the 
same structure as the initial conditions, i.e.~unperturbed by a planet.  
Regions with $|\phi|\geq\phi\sub{max}$ are 
approximated as ``strips'' extending from $\phi\sub{max}$ to $\pi/2$ 
and $-\phi\sub{max}$ to $-\pi/2$ .  
The surfaces interior to $r\sub{min}$ and exterior to $r\sub{max}$ are 
approximated as power laws ($h \propto r^{a}$) 
and integrated over as a series of 
infinite-length strips extending in the $\phi$ direction.   

\subsection{Initial Conditions}

Since we will be calculating the perturbed disk iteratively to 
take into account feedback between temperature and density perturbations, 
the initial disk itself must be stable to iterative feedback. 

We calculate the initial conditions for a slice in the middle of 
a box twice the size in $\phi$-space.  That is, the slice is at 
$\phi=0$ in a box ranging from $-\phi\sub{max}$ to $\phi\sub{max}$ 
with twice the number of grid points in $\phi$, and the 
same range of $r$ and $z$.  Outside the range of $\phi$, 
the surface of the disk is approximated as strips from $\phi\sub{max}$ 
to $\pi/2$.  Outside the range of $r$, the surface is approximated 
as strips with infinite length in the $\phi$ direction, 
with the height $z$ varying as a power law in $r$, fitted to the 
8 points at the ends of the box.  

The unperturbed (i.e.~without planet) structure of the disk 
is calculated iteratively as described above, but since azimuthal 
symmetry is assumed, this computation is significantly shorter than 
the computation of the perturbed disk structure.  Thus, we 
iterate until the rms change in disk height is less than 
$10^{-6}$, typically less than about 200 iterations. 

\begin{figure*}
\plottwo{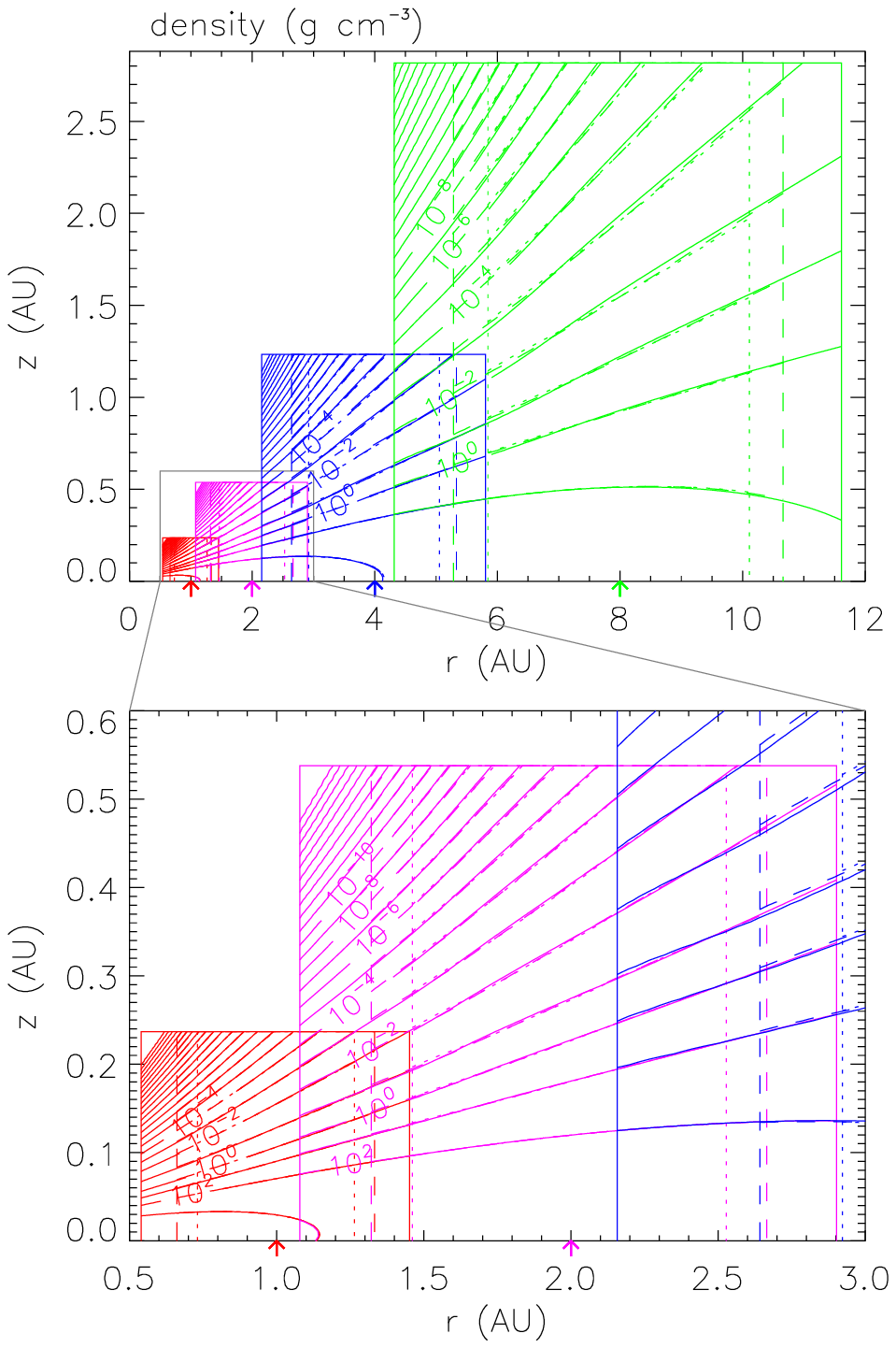}{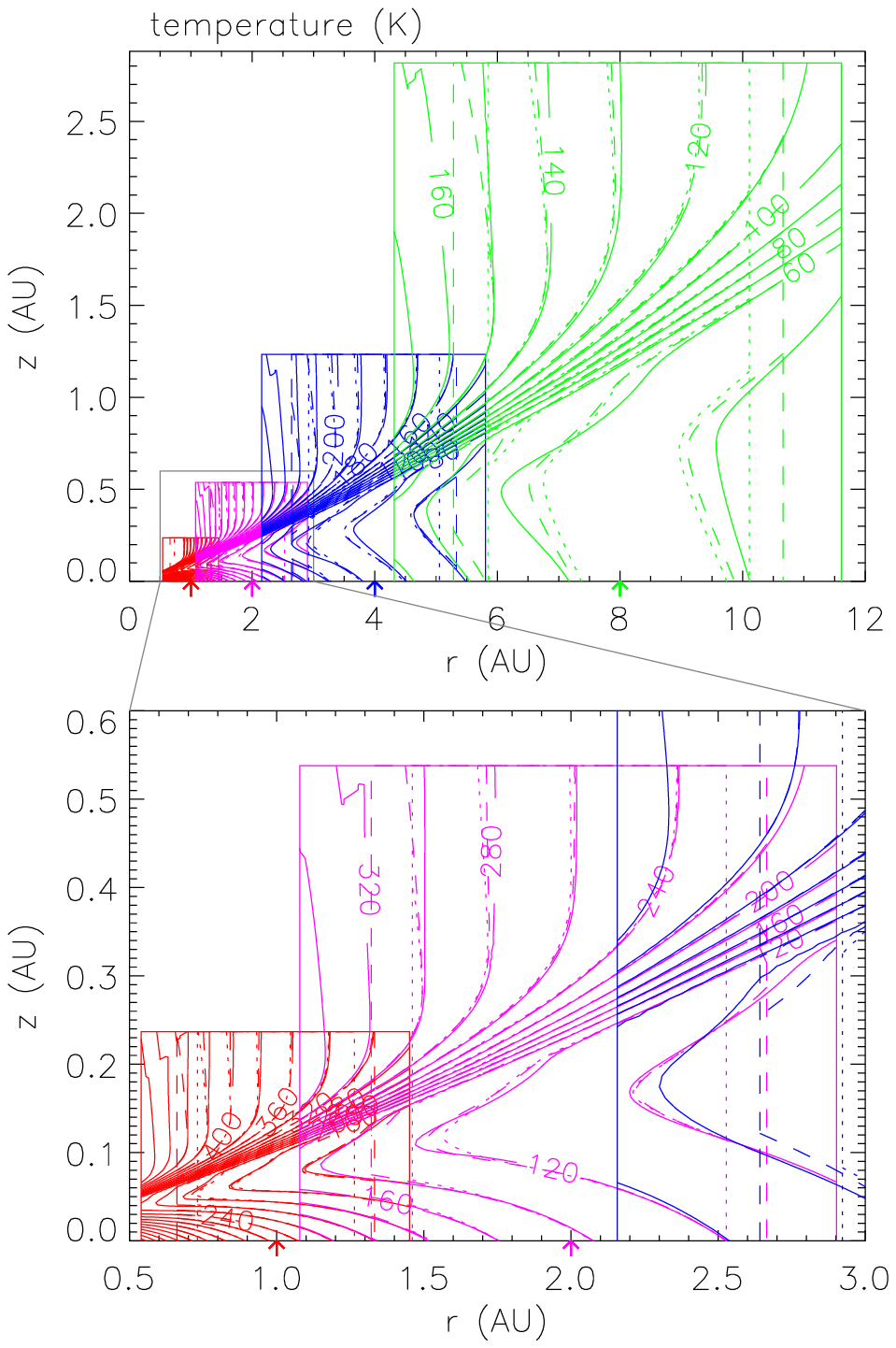}
\caption{\label{initcond}Initial density (left) and temperature (right)
profiles for the planet masses and distances studied in this paper.  
The lower set of plots are blow-ups of the area delimited by 
grey lines.  
Positions of planets at 1 (red), 2 (magenta), 4 (blue), and 8 (green) AU 
are indicated by arrows along the $r$-axis.  The density and temperature 
contours are indicated by dotted, dashed, and solid curves for 
10 M$_{\earth}$, 20 M$_{\earth}$, and 50 M$_{\earth}$ planets, respectively.  
\notetoeditor{This figure should span two columns}
}
\end{figure*}

We examine planets at $a=1$, 2, 4, and 8 AU with 
masses of 10, 20, and 50 M$_{\earth}$.  The sizes and positions of the 
simulation boxes for each of these parameters is indicated 
in \figref{initcond}.  
The limits of the simulation boxes in the $r$ direction are indicated 
by straight vertical lines: dotted, dashed, and solid for 
10 M$_{\earth}$, 20 M$_{\earth}$, and 50 M$_{\earth}$ planets, respectively.  
The boxes span from $r\sub{min}=a-12.5r\sub{Hill}$ to 
$r\sub{max}=a+12.5r\sub{Hill}$ in $r$, 
and from $0$ to $z\sub{max}\sim 2z_s$ where $z_s$ is the 
height of the unperturbed disk surface at $r=a$.  

The left set of plots in \figref{initcond} show the density 
contours of the initial conditions for each set of planet parameters, 
dotted, dashed, and solid curves for 
10 M$_{\earth}$, 20 M$_{\earth}$, and 50 M$_{\earth}$ planets, respectively.
The lower plot is a blow-up of the grey-outlined box in the upper plot.  
The density contours show a good amount of overlap over the range of 
parameter space.  

The right set of plots in \figref{initcond} is similar to the left set,
except it shows temperature rather than density.  The temperature contours 
do not overlap as well as the density contours, particularly at 
larger radii.  At the planets positions, the temperature contour converge 
fairly well, but tend to diverge at the boundaries of the simulations.  
This results from the sensitivity of the temperature structure to the 
density structure of the disk.  At small distances, the temperatures 
are domniated by viscous heating, so the shape of the surface is less 
relevant.  However, viscous heating drops off more rapidly than 
stellar irradiation heating, so at large distances, slight 
deviations in the calculation of the surface can lead to changes of 
several degrees in the vertical temperature structure of the disk.  
The discrepancies in temperature  may be overcome by increasing the 
size of the simulation box.  However, this leads to a corresponding 
increase in the computation time.  Since the difference in temperatures 
tends to be on the order of just a few degrees and since we are interested 
in the changes in disk structure due to the influence of a planet 
in the disk, at this time we will adopt the initial conditions as 
presented here.  

\section{Results}\label{results}

The planet is instantaneously inserted into the initial conditions and the 
resulting density and temperature perturbations are calculated.  The 
density and temperature are then iteratively recalculated under the 
assumption of hydrostatic equilibrium for 10 iterations.  Because the 
planet is instantaneously inserted into the disk and the assumption 
of hydrostatic equilibrium, 
there is no time scale associated with each iteration.  
Rather, the goal is to iterate until steady state is achieved.  
When evolution is discussed in this paper in the context of our disk-planet 
models, we refer to the changes in disk structure over successive 
iterations rather than a time sequence.  

\subsection{Self-Consistency: Effect of Iteration}

\begin{figure*}
\plottwo{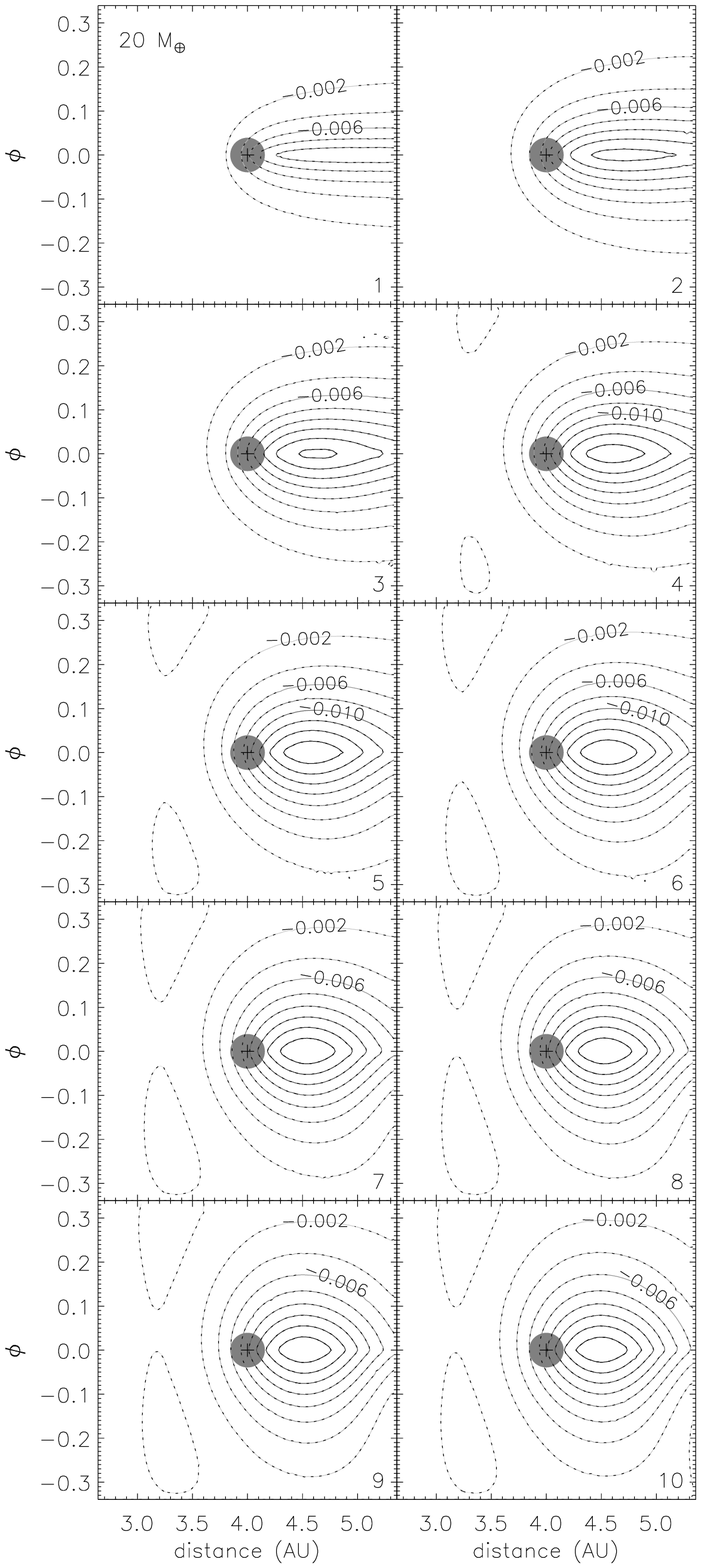}{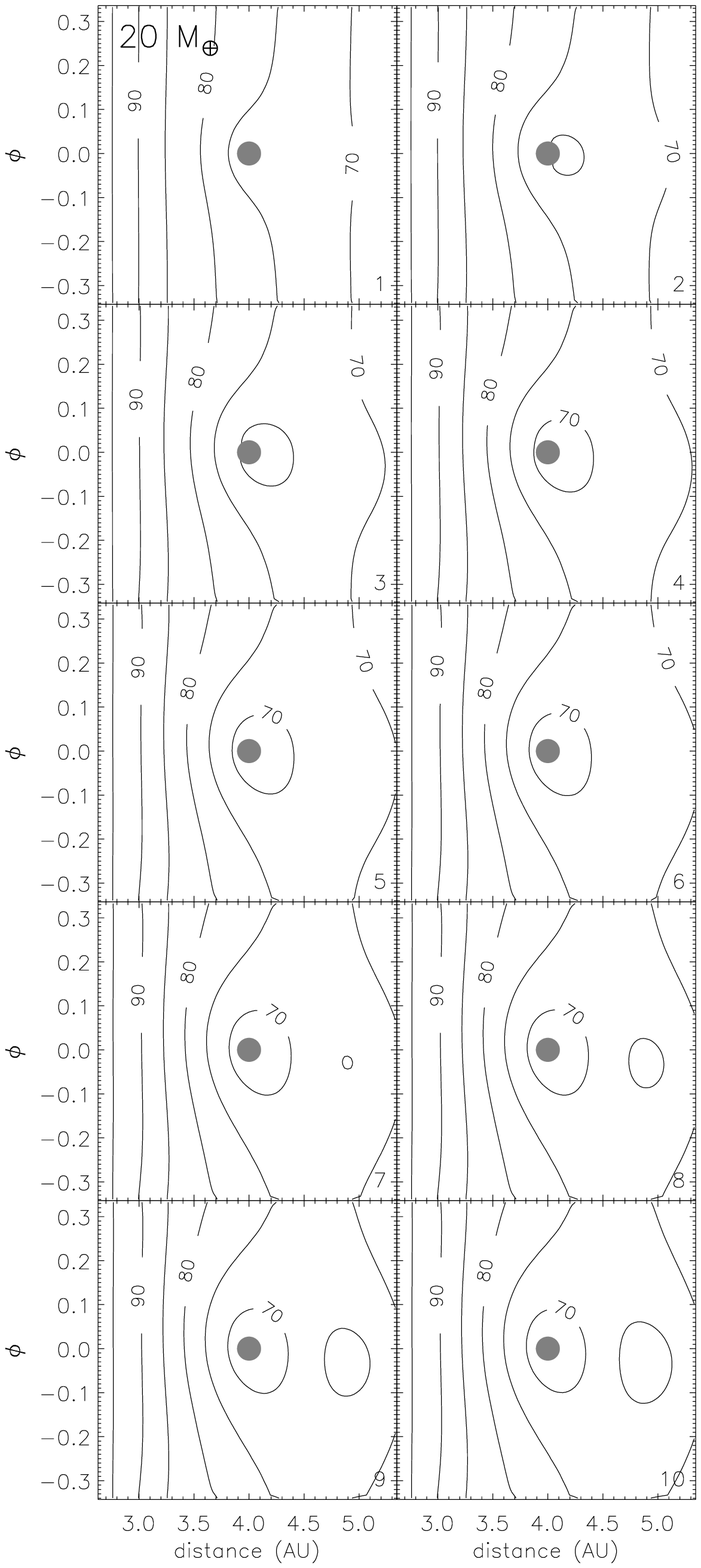}
\caption{\label{timeevol} Evolution of the surface (left) and 
photosphere temperature (right) for a 20 earth mass planet at 4 AU. 
The iteration number is indicated at the lower right in each 
individual plot.  (left) The deviation of the surface from the initial 
conditions is displayed as contours, in units of AU.  The 
grey circle is a Hill radius in size, centered at the planet's location.  
(right) Contour plots of temperatures in the photosphere, in kelvins.  
}
\notetoeditor{This figure should span two columns}
\end{figure*}

The iterative calculation was implemented to achieve self-consistency
between the temperature and density profiles of the disk under
hydrostatic equilibrium.  This self-consistency proves to 
be important to determining the structure of the disk. 
As an example, we examine the case of a 20 M$_{\earth}$ planet 
at 4 AU.  In \figref{timeevol} we show the evolution of the 
surface perturbation (left) 
and corresponding evolution of the temperature in the 
photosphere of the disk (right).    
The sequence of plots goes from left to right, top to bottom.  
The grey shaded circle shows the projected size of the Hill
sphere, looked down on the disk along the $z$-axis.  

Contours in the left panel show the fractional 
deviation of the surface at the given iteration number from the 
initial conditions, spaced at intervals of $0.002$.  
Contours in the right panel show temperatures in the photosphere 
of the disk, where the photosphere is defined to be where the 
optical depth using the Rosseland mean opacity 
integrated along the $z$-axis from $z\rightarrow\infty$ 
toward $z=0$ equals $2/3$.  Although the perturbation is quite subtle, 
less than $2\%$, the change in the temperature at the photosphere is 
dramatic by comparison, on the order of $10\%$ or more.  

With successive iterations, the perturbation to the surface 
grows in area and deepens.  This is seen in both the surface contours 
and photospheric temperatures.  This is because the region in shadow cools 
and compresses, deepening the shadow further still.  
Beyond the shadow, where the disk rises above the shadow, material 
heats and expands, causing this material to rise still further.  
The growth of the perturbation is limited both because of the differential 
rotation of disk material and because of the intrinsic 
radial temperature variation of the disk.  The differential rotation of 
the disk means that material more radially distant from the planet 
moves past the planet faster, so it has less time to heat or cool as 
it passes by the planet.  The intrinsic temperature structure of the disk 
means that inwards (outwards) of the planet, 
the disk is hotter (cooler), limiting the growth of the shadow 
(brightened region) in that direction.   
Iterations $8 - 10$ show very little difference between each other, indicating 
that the density and temperatures have reached self-consistency.  

\begin{figure}
\plotone{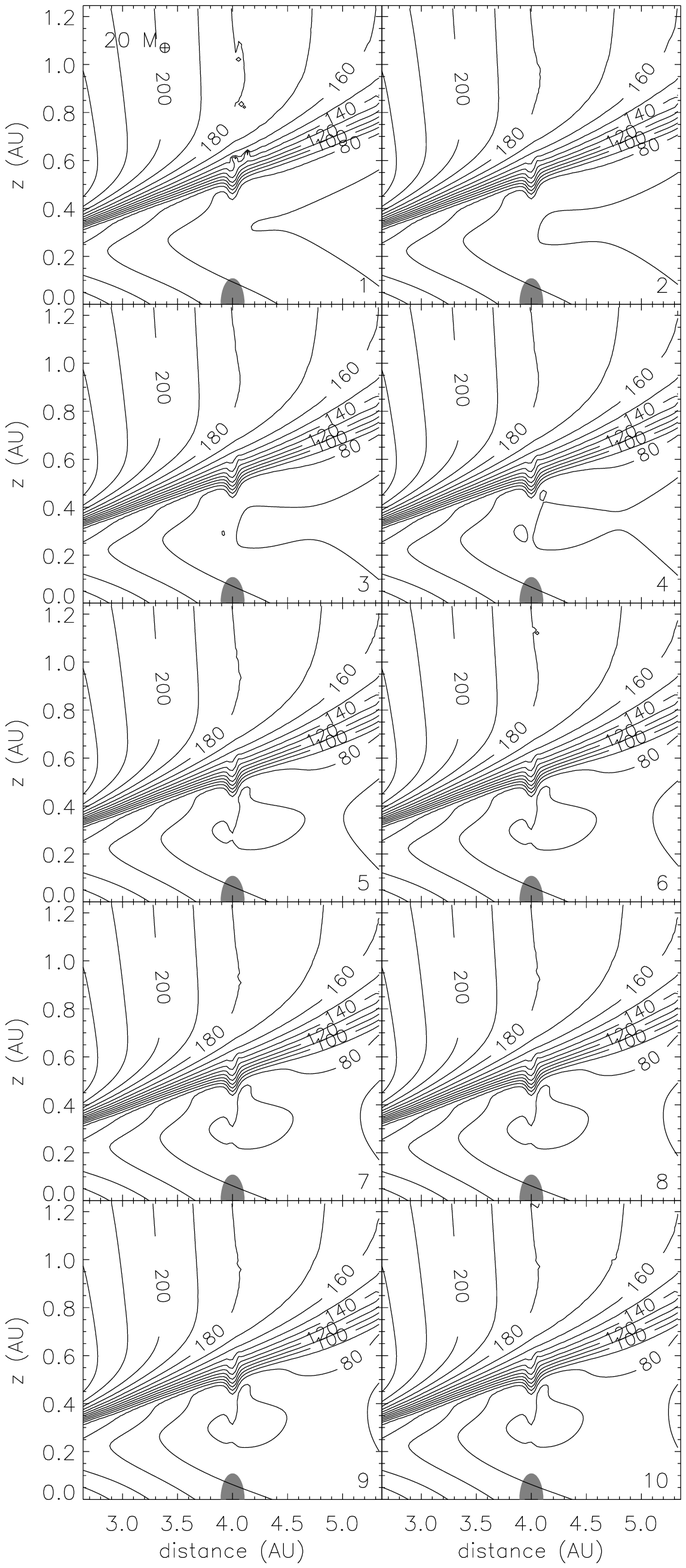}
\caption{\label{sliceevol}Evolution of temperatures 
in a cross-section of the disk at $\phi=0$ for a 20 earth mass 
planet at 4 AU. 
The Hill radius and position of the planet are indicated by the grey 
half-ellipse.  
The temperatures are indicated by contours, with the 
iteration number indicated at the lower right in each 
individual plot.  
}
\end{figure}

\figref{sliceevol} shows another view of the effect of iteration 
on the temperature structure of the disk.  Here we plot the temperature 
cross-section of the disk at $\phi=0$, the location of the planet.  
The size of the Hill sphere is represented by the grey ellipse 
(the $r$ and $z$-axes are not scaled to each other). 
The surface layers are relatively unaffected by the the changing density 
structure, largely because this region is optically thin, and the 
dependence of temperature on angle of incidence is small.  
The midplane layer is relatively unaffected as well.  This is due to 
several effects: viscous heating is greatest at the midplane; 
the angular sizes of the shadowed/brightened regions are small; and 
the disk is optically thick.  
The intermediate layers, change in temperature structure 
quite a bit.  The development of a cooled region directly above 
the planet and the heated region outward from the planet 
are quite clearly evident.  

\begin{figure}
\plotone{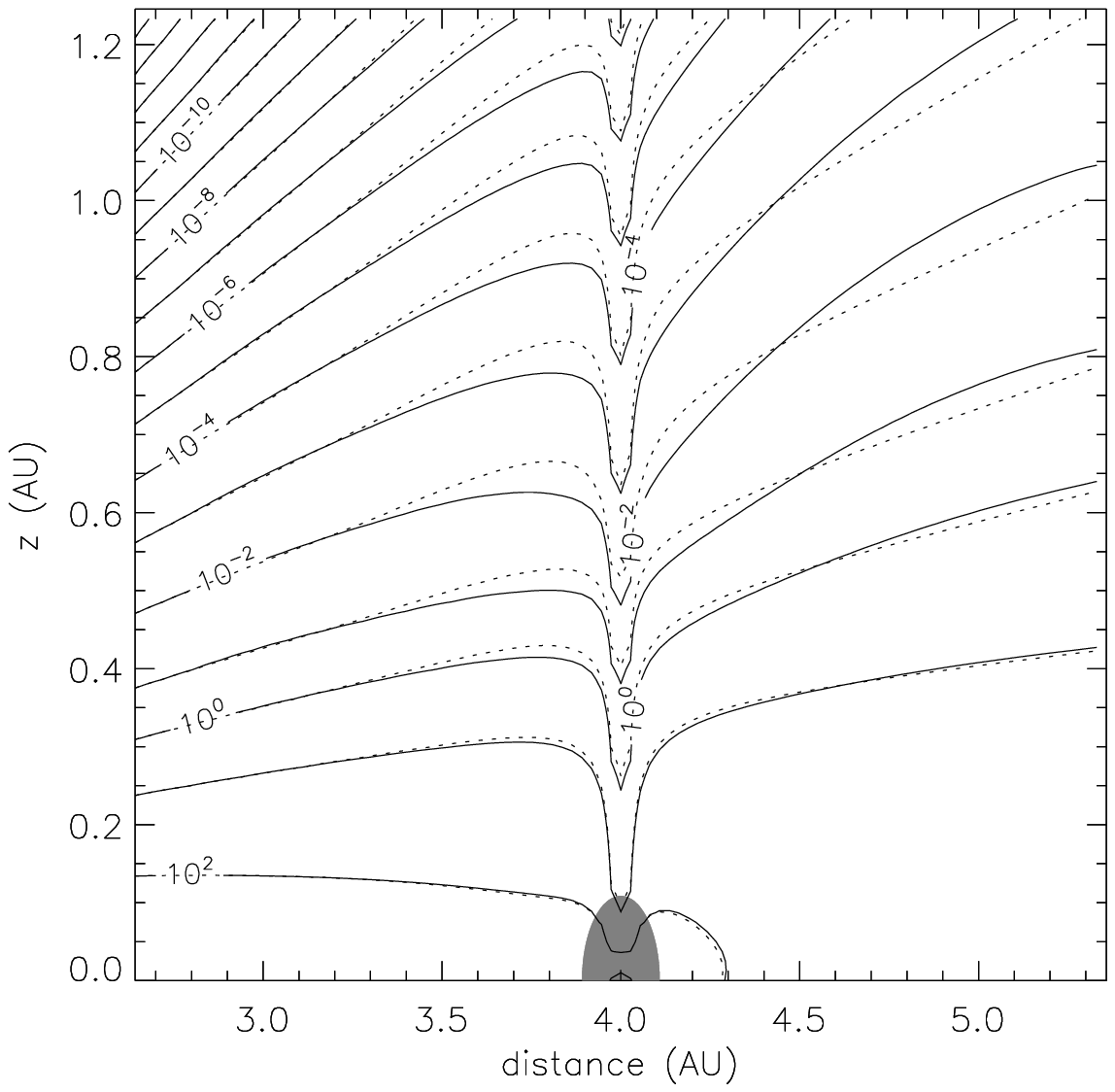}
\caption{\label{denevol}Evolution of densities 
in a cross-section of the disk at $\phi=0$ for a 20 earth mass 
planet at 4 AU. 
The Hill radius and position of the planet are indicated by the grey 
half-ellipse.  
The density varies subtly with successive iterations, so 
only density contours after the first iteration (dotted lines) 
and tenth iteration (solid lines) are shown. 
}
\end{figure}

Meanwhile, the density structure changes only subtly.  
In \figref{denevol}, we plot the density contours for the $\phi=0$ slice 
of the simulation box after 1 (dotted lines) and 10 
(solid lines) iterations.  Over the 10 iterations, the density contours
do change, but not as dramatically as the temperature perturbations.  
This emphasizes the importance of the detailed density structure 
to the calculation of radiative transfer in the disk.  

\subsection{Variations with Planet Mass and Distance}

\begin{figure*}
\plotone{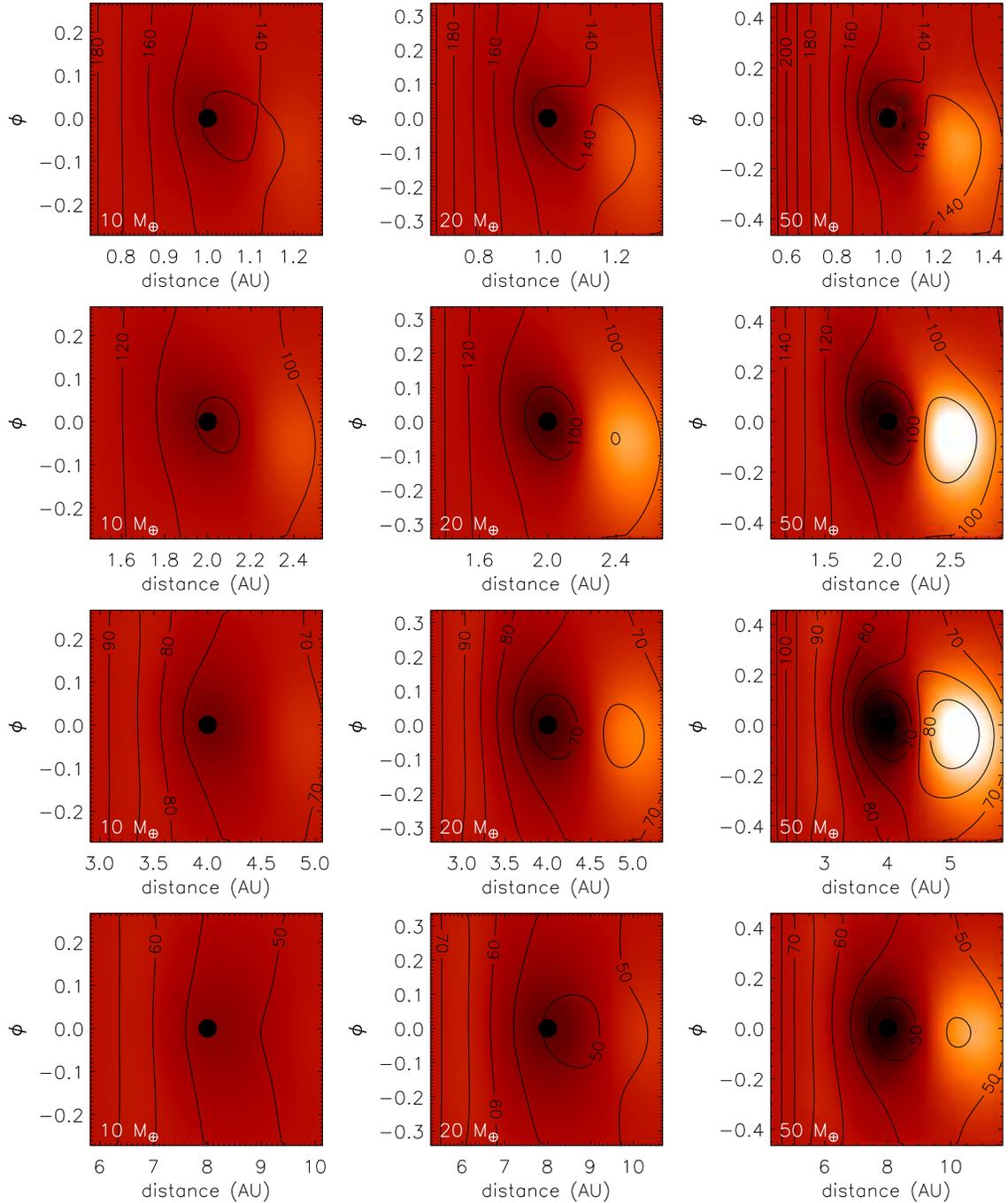}
\caption[photospheres.eps]{\label{photospheres}Temperatures in the photospheres of disk 
models with planets after 10 iterations.
From top to bottom: planets at 1 AU, 2 AU, 4 AU, and 8 AU.  
From left to right: planets with mass 10 M$_{\earth}$, 20 M$_{\earth}$, 
and 50 M$_{\earth}$.
The region above the Hill sphere is blacked out.  
Solid contours indicate temperatures in kelvins.  
The shading shows the fractional deviation in temperature from the 
initial disk model, 
so that black is $\Delta T/T=-0.2$ and white is $\Delta T/T=+0.2$
}
\end{figure*}

We investigate the changes to the disk temperature structure 
as planet mass and distance vary. We examine planets  
with masses of 10, 20 and 50  M$_{\earth}$
at 1, 2, 4, and 8 AU.  
\figref{photospheres} summarizes the changes in photosphere 
tempeartures for this suite of planet parameters.  
The contours in each of these plots indicate the absolute 
temperatures, while the greyscale reflects the fractional temperature 
variation from the initial conditions, dark for cooling and 
lighter for heating.  In general, the cooled region is 
centered at the planet's position, while the headed region is 
outside the planet's orbit.  The asymmetry in the perturbations 
are a result of including the differential rotation of the disk 
and a finite time scale for cooling/heating as a parcel of gas 
passes through a shadow/brightening.  The sizes of the perturbations 
roughly scale both with planet mass and distance.  

\begin{figure*}
\plotone{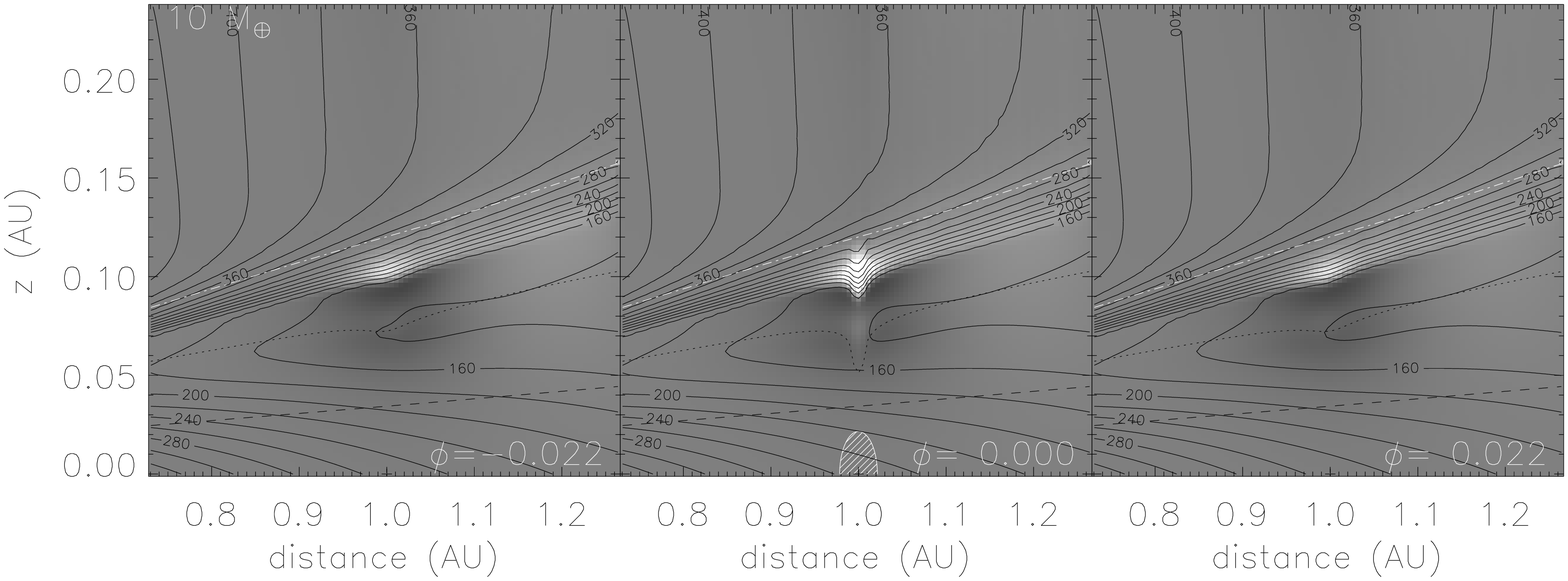}
\plotone{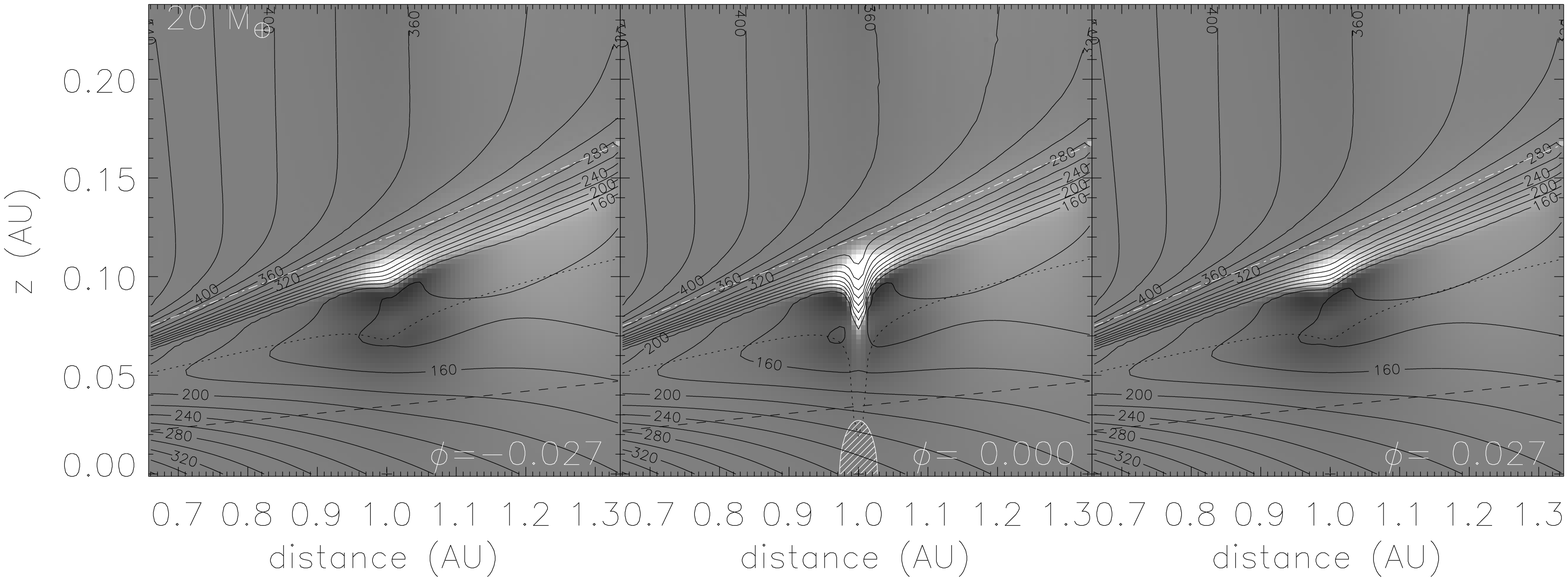}
\plotone{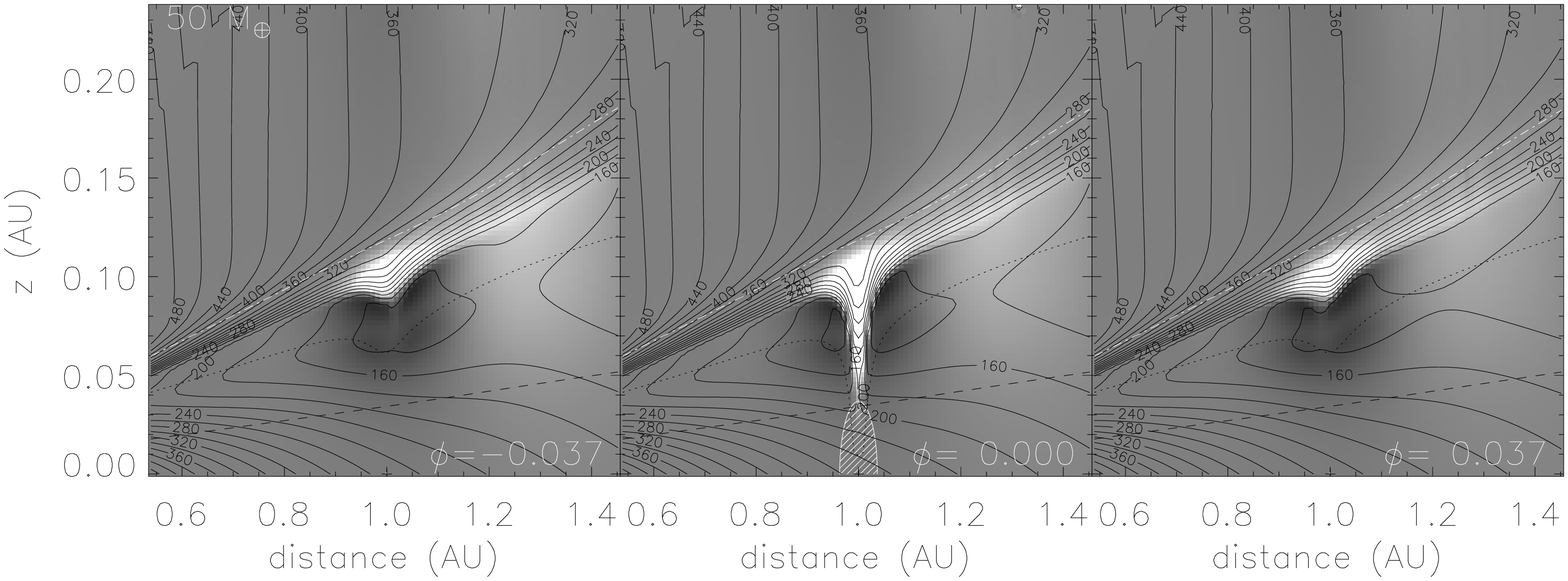}
\caption{\label{tempslice_a1}Temperature cross-sections 
of the disk in the vicinity of a planet at 1 AU from a 1 M$_{\sun}$ 
star at the indicated 
azimuthal angles, corresponding to (from left to right) 
1 Hill radius downstream 
of the planet, at the planet position, and 1 Hill radius 
upstream of the planet.  
The location and size of 
the Hill sphere are indicated by the white hashed area.  
From top to bottom, planet masses are 10 M$_{\earth}$, 20 M$_{\earth}$
and 50 M$_{\earth}$.  
The contours show absolute temperatures while the color scale show 
the fractional temperature difference from the initial 
disk model, with a range of $\Delta T/T \in [-0.2,+0.2]$.  
}
\end{figure*}

\begin{figure*}
\plotone{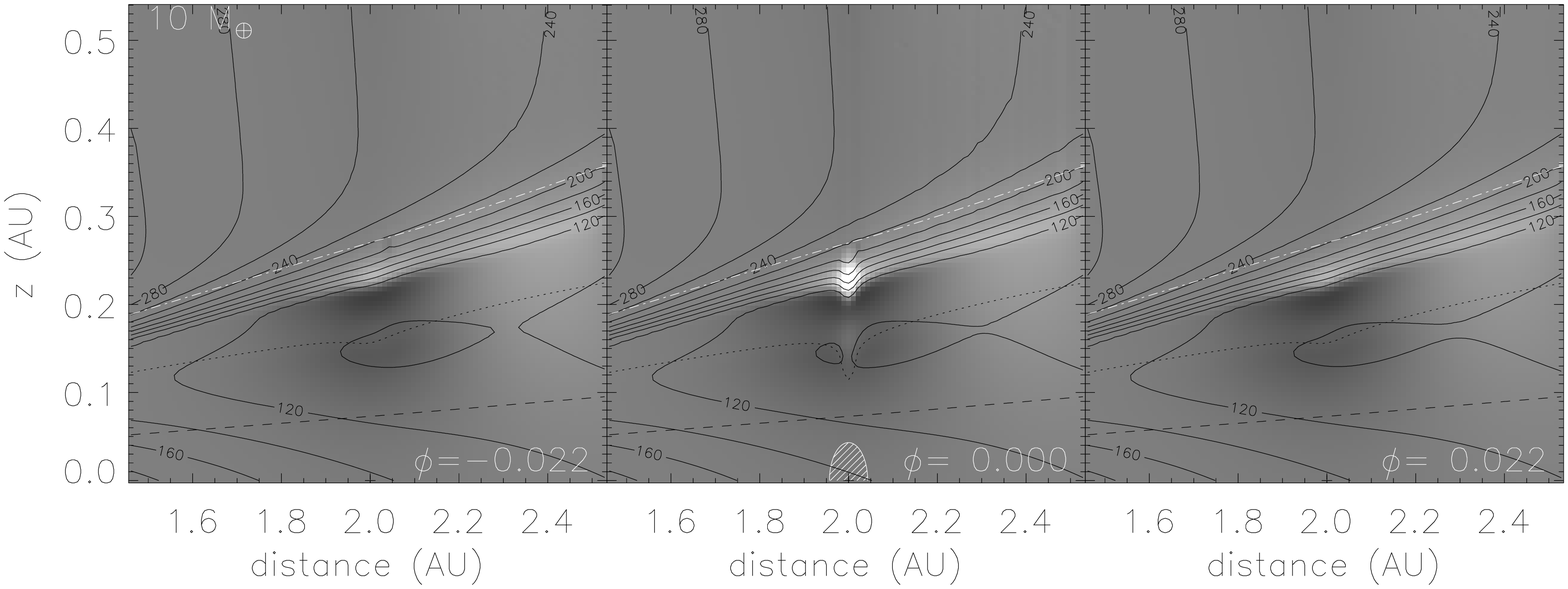}
\plotone{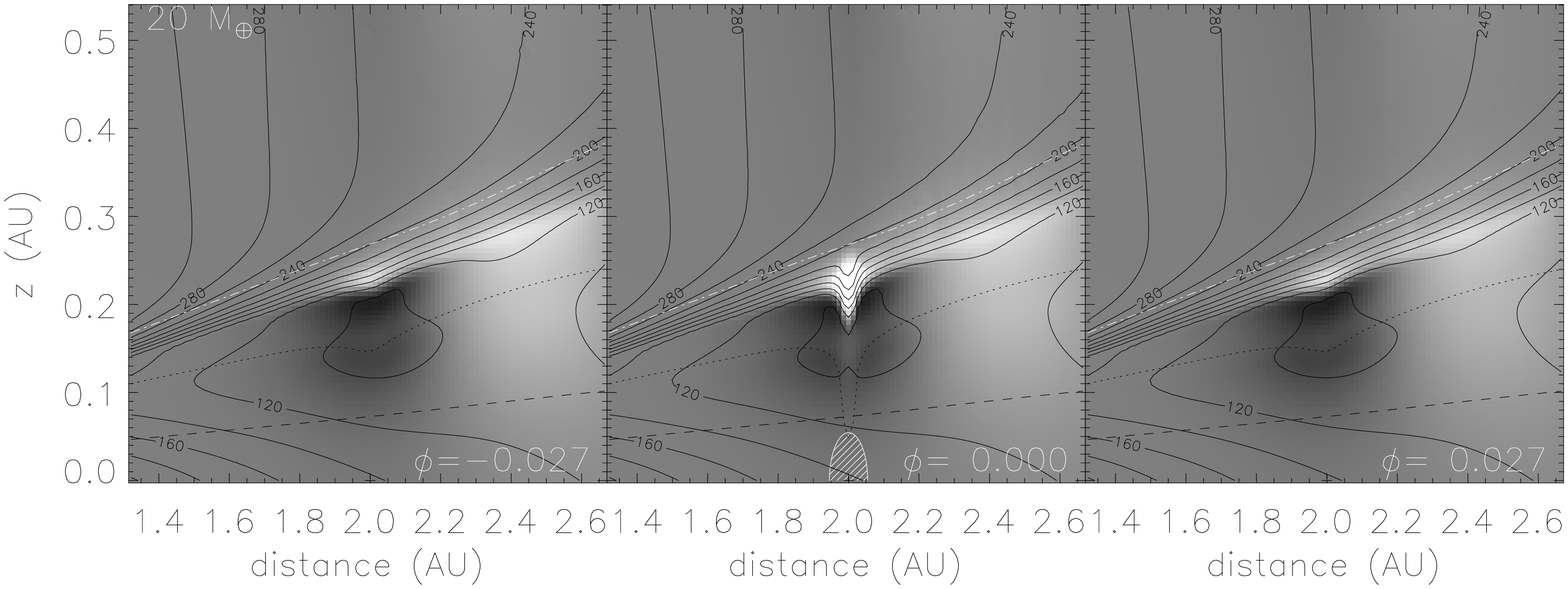}
\plotone{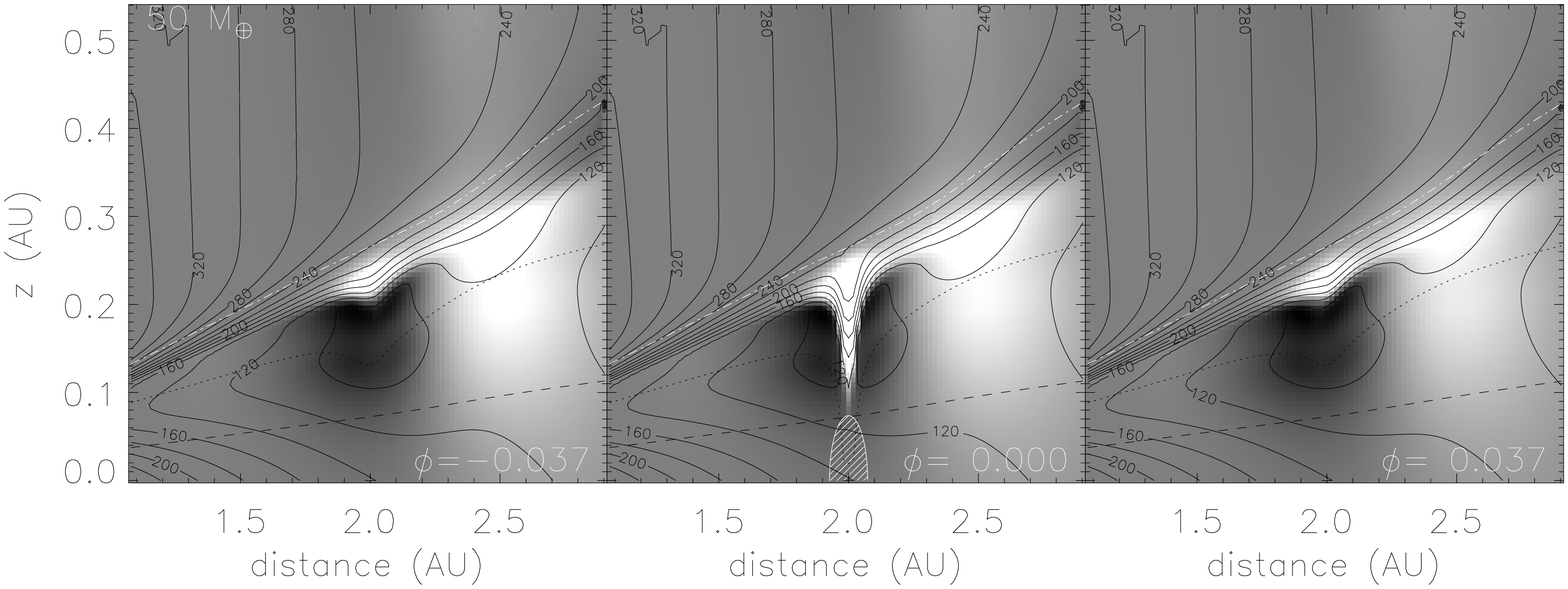}
\caption{\label{tempslice_a2}Same as \figref{tempslice_a1},
for planets at 2 AU.}
\end{figure*}

\begin{figure*}
\plotone{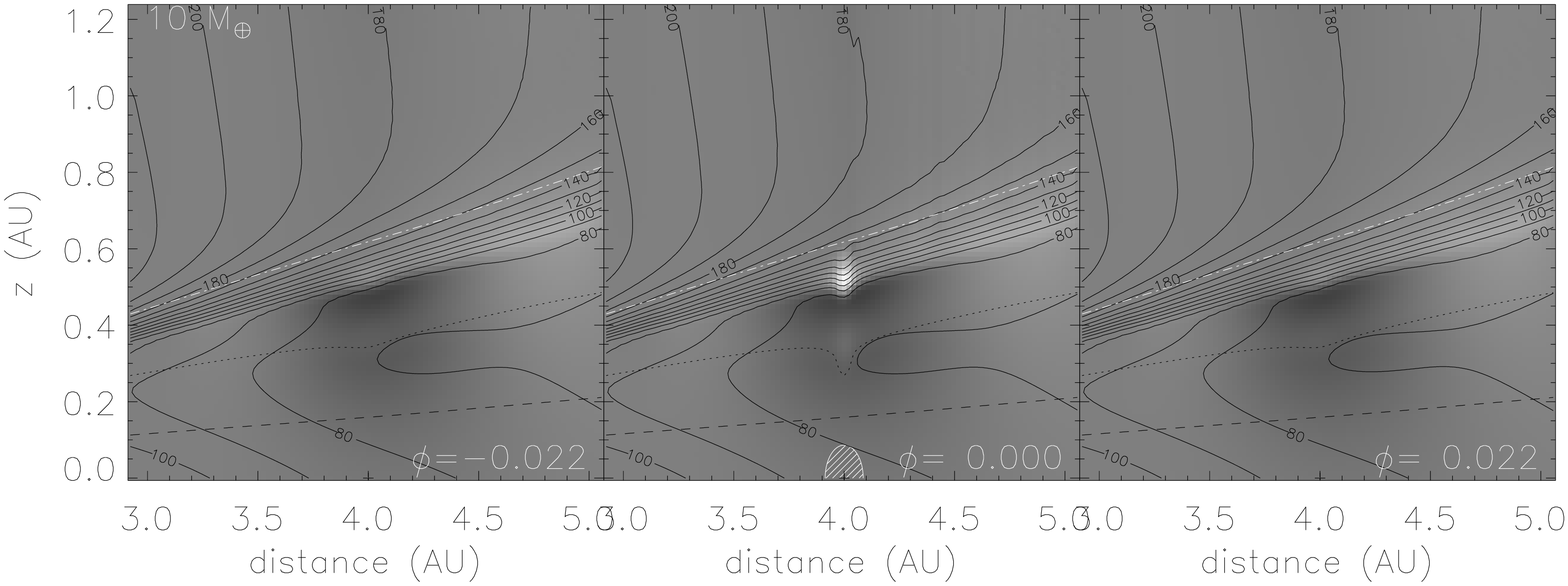}
\plotone{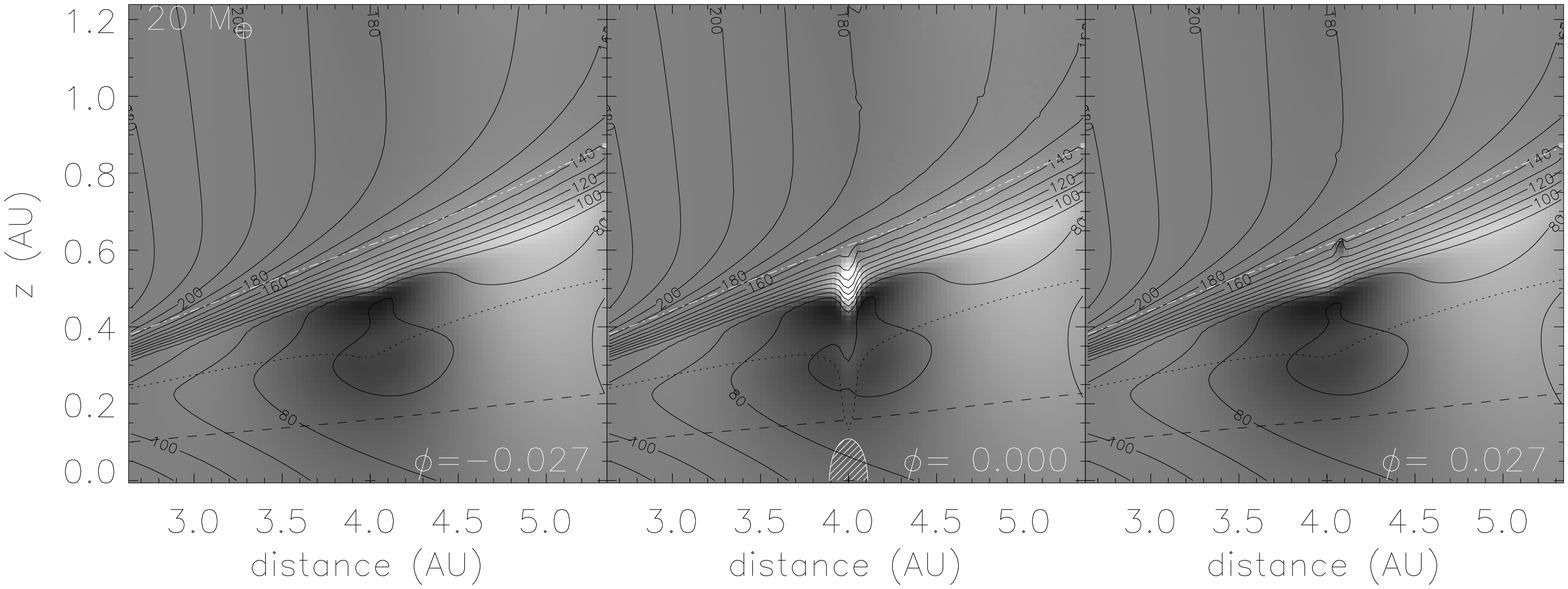}
\plotone{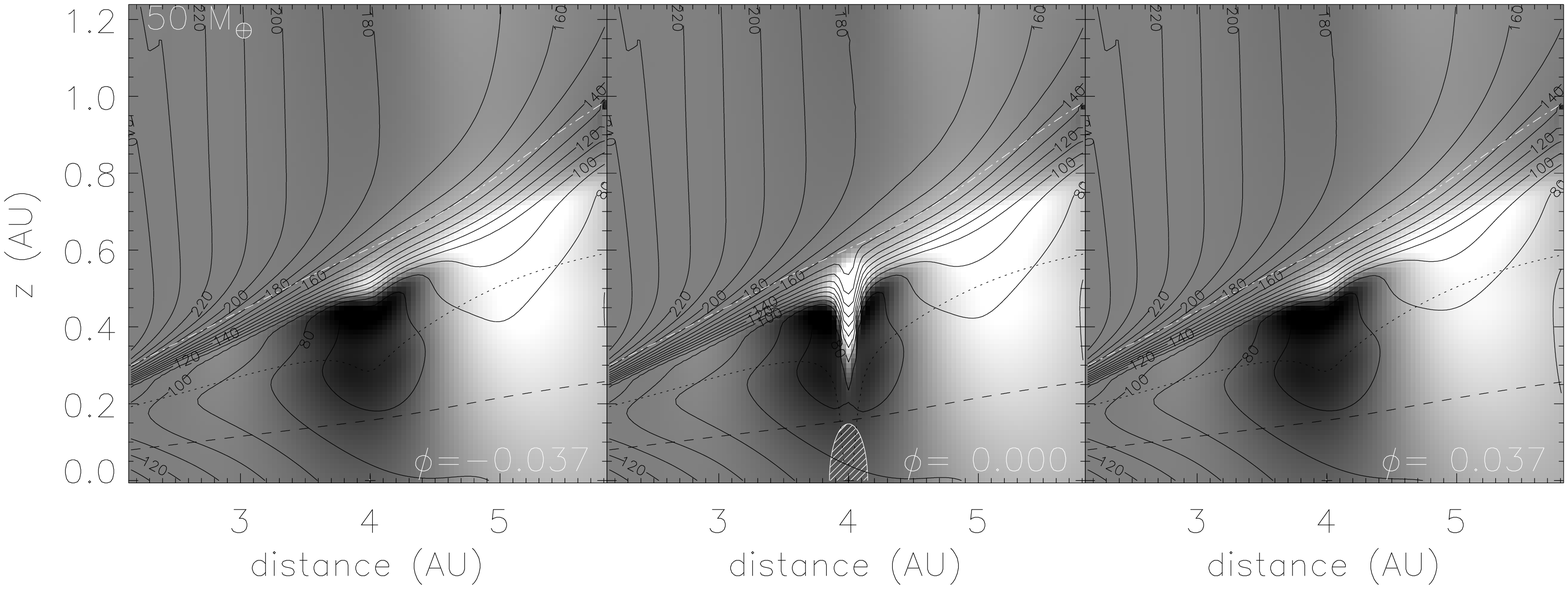}
\caption{\label{tempslice_a4}Same as \figref{tempslice_a1},
for planets at 4 AU.}
\end{figure*}

\begin{figure*}
\plotone{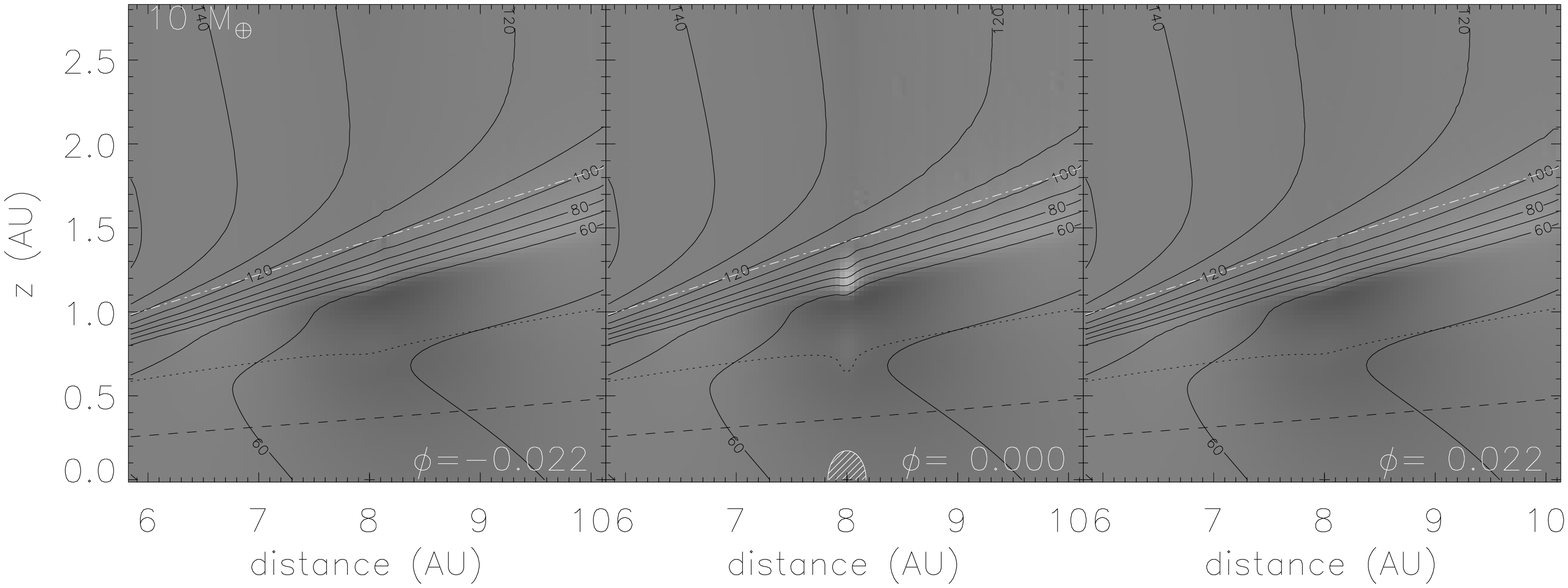}
\plotone{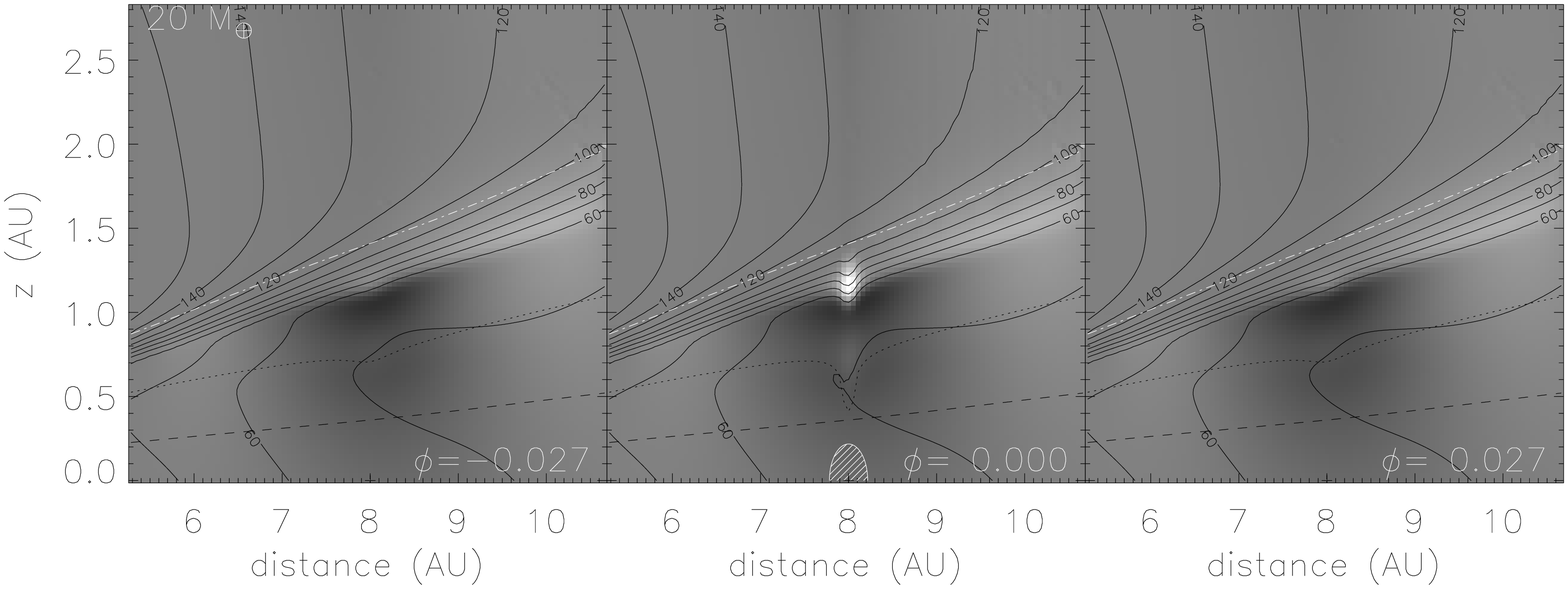}
\plotone{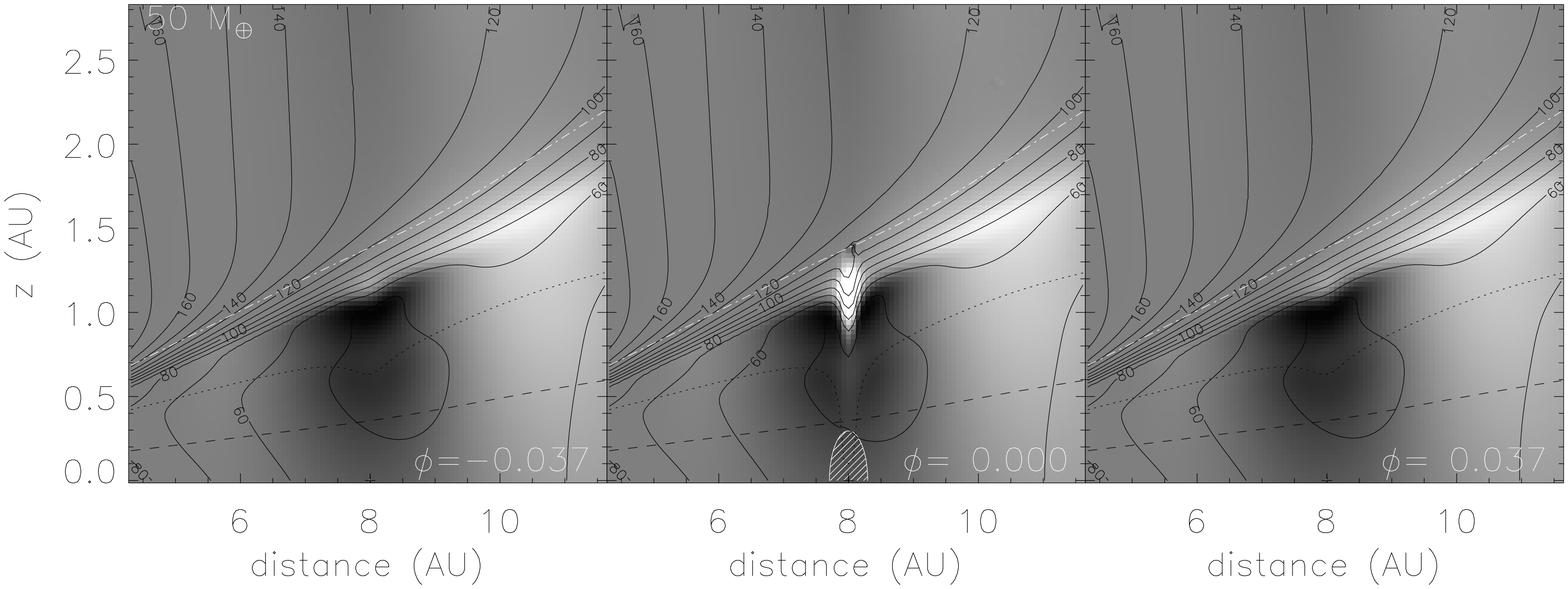}
\caption{\label{tempslice_a8}Same as \figref{tempslice_a1},
for planets at 8 AU.}
\end{figure*}

In Figs.~\ref{tempslice_a1}-\ref{tempslice_a8}, we show 
radial cross-sections of the temperature structure at 
(from left to right) 
1 Hill radius downstream from the planet, at the planet position, 
and 1 Hill radius upstream of the planet, for the planet parameters 
examined in this study.  The solid lines show the temperature 
contours, while the greyscale shows the amount of deviation from the 
initial conditions. 
Figs.~\ref{tempslice_a1}, \ref{tempslice_a2}, \ref{tempslice_a4}, 
and \ref{tempslice_a8} show planets at 1, 2, 4, and 8 AU, respectively.  
For reference, 
the locations of the photospheres are indicated as black dotted lines.  
Also shown are the surface of the disk (white dot-dashed line) 
and thermal scale height (black dashed line).  
Both the photosphere and surface are multiple scale 
heights above the disk, except just above the Hill sphere.  
Temperature perturbations tend to be greatest in the upper layers 
of the disk, between the photosphere and the surface.  
Again, this is because of viscous heating at the midplane as 
well as optical depth effects.

\section{Discussion}\label{discussion}

\subsection{Comparison to Previous Models}\label{prevmod}

We now compare our results to those of Papers I and II.  
The results presented there were for planets around a star 
of 0.5 M$_{\sun}$, while those presented here are for a 1 M$_{\sun}$ 
star.  In order to make a fair comparison, we have recomputed 
the model presented in Papers I and II for the stellar parameters 
used in this paper.  
We will henceforth refer to this set of models as the J-CS model.

\begin{figure}
\includegraphics[width=3.5in]{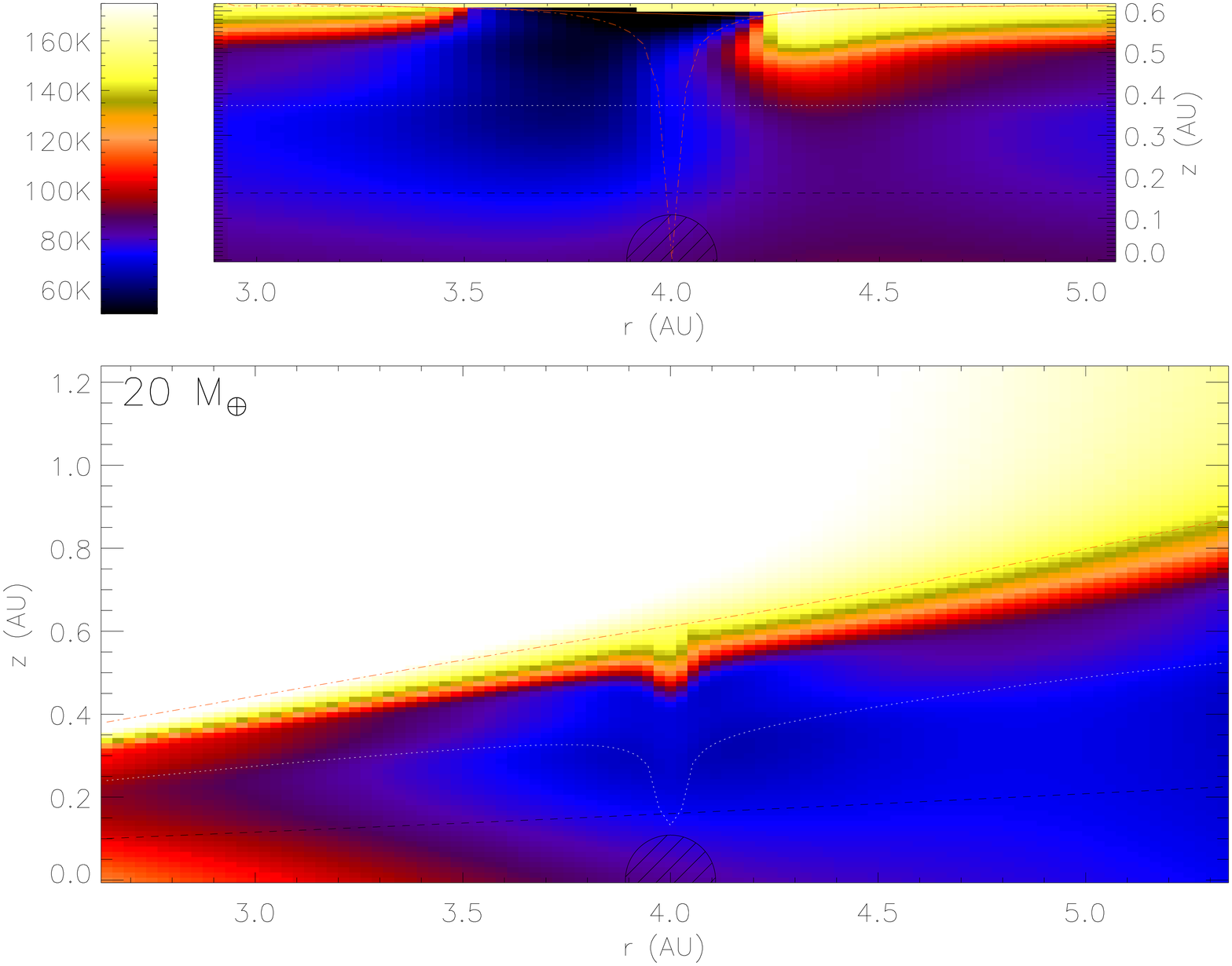}
\caption{\label{compare_a4p20}The thermal structure of the disk in 
the vicinity of a 20 M$_{\earth}$ planet at 4 AU from a 1 M$_{\sun}$ 
star, as calculated for (top) J-CS and (bottom) 
this paper.  The temperatures are scaled to the same 
colors, and the $r$ and $z$ axis have the same scaling.  
The hashed semi-circle shows the size and location of the Hill sphere.  
The photosphere is indicated by white dotted lines 
and the thermal scale height is indicated by black dashed lines.  
In the top plot, the red dot-dashed line shows the isodensity contour 
equal to the density at the disk surface.  Regions below the solid 
red line in the top plot are considered to be in shadow.  
In the bottom plot, the red dot-dashed line is the surface as 
calculated from line-of-sight optical depth to the star.  
}
\end{figure}

In \figref{compare_a4p20}, we compare the results from this paper to 
J-CS for the case of a 20 M$_{\earth}$ 
planet at 4 AU from its host star.  The top plot shows the 
radial temperature cross section through the planet's position 
for J-CS and the bottom shows the same cross-section 
for the new model.  In the J-CS model, the disk is treated as 
locally plane-parallel, whereas in the new model it is not.  
This is reflected in the shape of the temperature structures 
of the two models.  Another difference is that the surface 
was calculated as the contour of constant density in J-CS.  
This contour is shown as a red dot-dashed line.  When this contour 
is shadowed, the surface is instead taken to be the line of 
sight to the star, indicated by the solid red line.  In contrast, 
the new model calculates the surface to be where the optical 
depth to the star using the Planck mean extinction, $\chi_P^*$, 
is $2/3$.  

\begin{figure}[b]
\hspace{-0.05in}
\includegraphics[width=1.7in,bb=0 10 176 170]{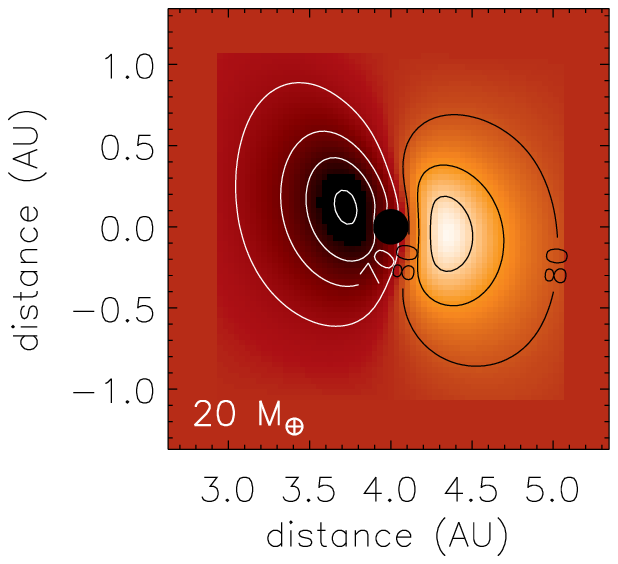}
\hspace{-0.1in}
\includegraphics[width=1.7in,bb=0 10 176 170]{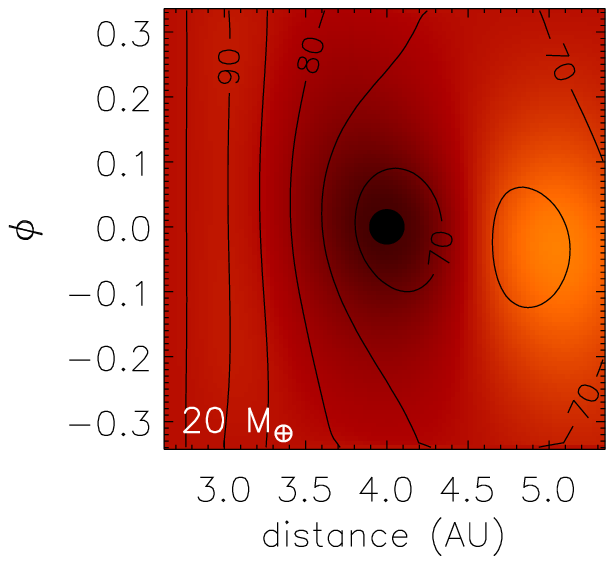}
\caption{\label{oldplot}
Temperature perturbations in the disks photosphere above an embedded planet, 
(left) as calculated in the J-CS model and 
(right) using the methods presented in this paper.  
Figures are scaled to the same spatial size, and colors represent 
fractional deviation from the unperturbed disk temperature.  
Contours show temperatures.  Since the J-CS calculation assumed 
a plane-parallel disk, the temperature contours trace the color 
variations.  In the calculation presented in this paper, 
the temperatures are dominated by the radial variation of the 
unperturbed disk.}
\end{figure}

In J-CS, the temperature perturbations are more dramatic 
than in the new model.  This can also be seen in \figref{oldplot}, 
which shows the temperatures in the photosphere compared 
side-by-side.  The plots show temperature contours and fractional
temperature variations, with the older model represented on the left. 
The right plot in this figure is the same as the one in 
\figref{photospheres}.  The spatial sizes and color scalings are the 
same between the two plots.  It is evident from both 
Figs.~\ref{compare_a4p20} and \ref{oldplot} that the temperature 
perturbations are much greater in J-CS.  The reason 
for this is primarily because of the way in which the surface was 
calculated.  As noted in Paper I, the disk temperature structure is 
quite sensitive to $\mu$, the cosine of 
the angle of incidence of stellar irradition 
at the surface.  In other words, the slope of the surface determines 
the amount of heating from stellar irradiation.  
Small deviations at the surface can lead to 
large temperature perturbations at the photosphere of the disk.  

In the JC-S model, $\mu\rightarrow 0$ in the shadowed region, which 
means that zero flux was propagated from shadowed surface elements 
to disk material below them.  When the density contour 
rose back above the shadow, the transition from shadow to 
the illuminated region was quite sharp, with relatively large 
values of $\mu$ yielding large amounts of heating.  
In the new model, since the surface is calculated by integrating 
the optical depth along lines of sight to the star, the transitions 
are smoother and values of $\mu$ do not vary as greatly.  
Also, as shown in \eqref{mu}, we set a minimum value for $\mu$ 
based on the finite size of the star, so shadowing can not 
result in as significant cooling as seen in J-CS.  
Yet another effect is the radial variation in stellar flux, which 
was not taken into account in the plane-parallel J-CS model.  
The disk is hotter/cooler closer to/farther from the star, 
which offsets some of the effects of shadowing/illumination.  
Although the J-CS model was not iterative while the new model 
is, the changes in the calculation of the surface and $\mu$ 
are sufficient to diminish the temperature perturbations 
near the planet.  

Another difference between J-CS and this work is the locations 
of the heated and cooled regions.  In J-CS, these regions 
are nearly symmetric with respect to the planet position.  
The sizes and shapes of the heated and cooled regions are 
similar, with the cooled region inward of the planet and the heated 
region outward of it.  In the current work, The cooled region 
in centered on the planet's position with the heated region 
outward of the planet. 

\subsection{Magnitude of Temperature Variations}

In Paper II, we found a correlation between the maximum and 
minimum photosphere temperature variations and the 
Hill radius of the planet.  We now perform the same analysis 
on the new model.  

\begin{figure}[b]
\plotone{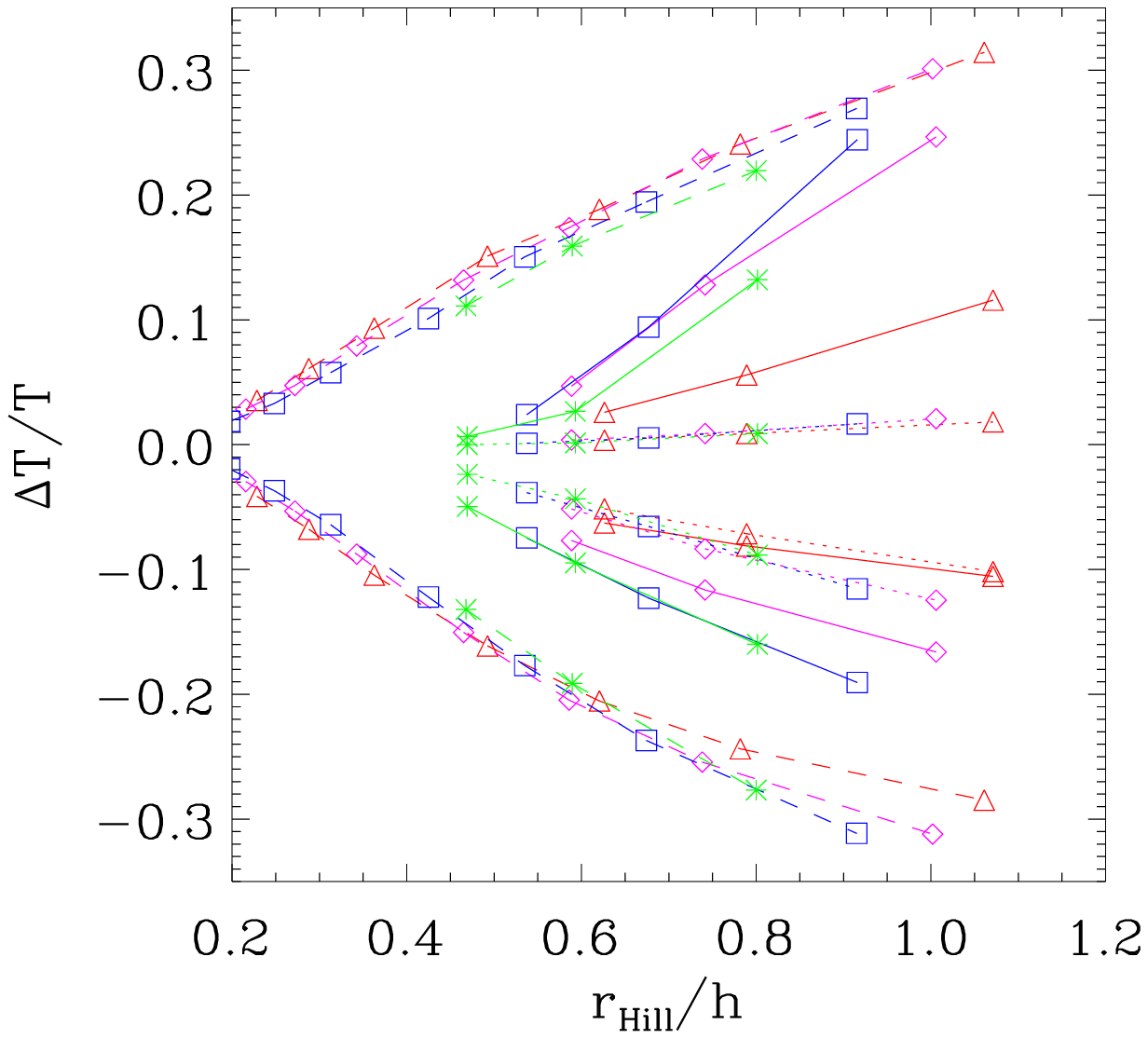}
\caption{\label{temp_vs_rhill}Maximum and minimum fractional temperature 
variation in the photosphere versus $r_{\mbox{\scriptsize Hill}}/h$.  
Symbols indicate planet distance as follows:
triangles at 1 AU, 
diamonds at 2 AU, 
squares at 4 AU, 
and asterisks at 8 AU.
Dashed lines show the J-CS model.  
Dotted lines show the current model, after only one iteration.  
Solid lines show the current model, after 10 iterations.  
}
\end{figure}

In \figref{temp_vs_rhill}, we plot the maximum and minimum fractional 
temperature variations in the photospheres versus the ratio of the 
Hill radius to thermal scale height of the disk ($r\sub{Hill}/h$).  
Points above/below $\Delta T/T =0$ represent maxima/minima.  
The symbols denote at what distance the planet is: 
triangles for 1 AU, 
diamonds for 2 AU, 
squares for 4 AU, 
and asterisks for 8 AU.
The results for J-CS are plotted connected by dashed lines.  
As noted in Paper II, the maxima and minima lie nearly on the 
same curves.  The upturn in minimum temperatures at smaller distances 
is due to the rise in viscous heating closer to the star, which 
sets the absolute temperature minimum.  

The points connected by dotted lines in \figref{temp_vs_rhill} 
are temperature maxima/minima in the photospheres 
after one iteration, i.e.~without 
recalculating the density given the new temperature structure.  
These points similarly lie on nearly the same curves, 
except showing a much smaller temperature perturbation 
than JC-S.  The reasons for the smaller temperature variation 
was detailed in \S\ref{prevmod}.  

The effect of including iterations can be seen by comparison 
to the points connected by solid lines in \figref{temp_vs_rhill}.  
In all cases, the fractional temperature variations increase.  
However, the amount of change is not consistent with planet 
mass or distance.  
At 1 AU (triangles), the change in temperature minima is nearly 
negligible.  This can be explained by viscous heating, which 
halts the growth of the cooled, shadowed region.  
The temperature maxima do increase, but not as much as at 
larger distances.  This is because the growth of the brightened region 
is coupled to the growth of the shadowed region: as the shadowed region 
grows, more disk material behind the shadow is exposed and illuminated.  
Since shadow growth is limited at 1 AU, so is the growth of the 
brightened region.  
Going from 2 to 4 to 8 AU, viscous heating ceases to be 
important.  The temperature maxima all lie close to the same curve at 
those distances, and the temperature minima at 4 and 8 AU 
lie nearly on the same curve in \figref{temp_vs_rhill}.  

For larger planets, those whose Hill radii approach the 
disk thermal scale height, the amount of heating in the 
current model approaches that seen in J-CS.  However, the 
temperature minima never drop that low.  This can 
be attributed to the minimum value of $\mu$ that has been 
adopted for the new model.  There is such upper limit 
that has been imposed however, hence the greater heating.

\subsection{Sizes of Thermally Perturbed Regions}

Although the variation in temperature appears to be less 
in the present model than that seen in J-CS, 
Figs.~\ref{compare_a4p20} and \ref{oldplot} suggest that 
the spatial scale of the perturbations might be larger.  
To check this, we calculate the areas of the photosphere that 
are heated or cooled above or below a certain threshold.  
We call these regions hot or cold ``spots.''  
Note that this is different from the definition of a 
``spot'' in Paper II, which considered regions heated or below 
170 K, in the context of ice formation and the snow line.  
Here, we consider the deviation in temperature from 
the unperturbed value, rather than above or below a 
fixed threshold.  A separate paper 
addressing the effect of these temperature perturbations on the 
snow line is addressed by \citet{HJCPodolakSasselov}. 

\begin{figure}
\plotone{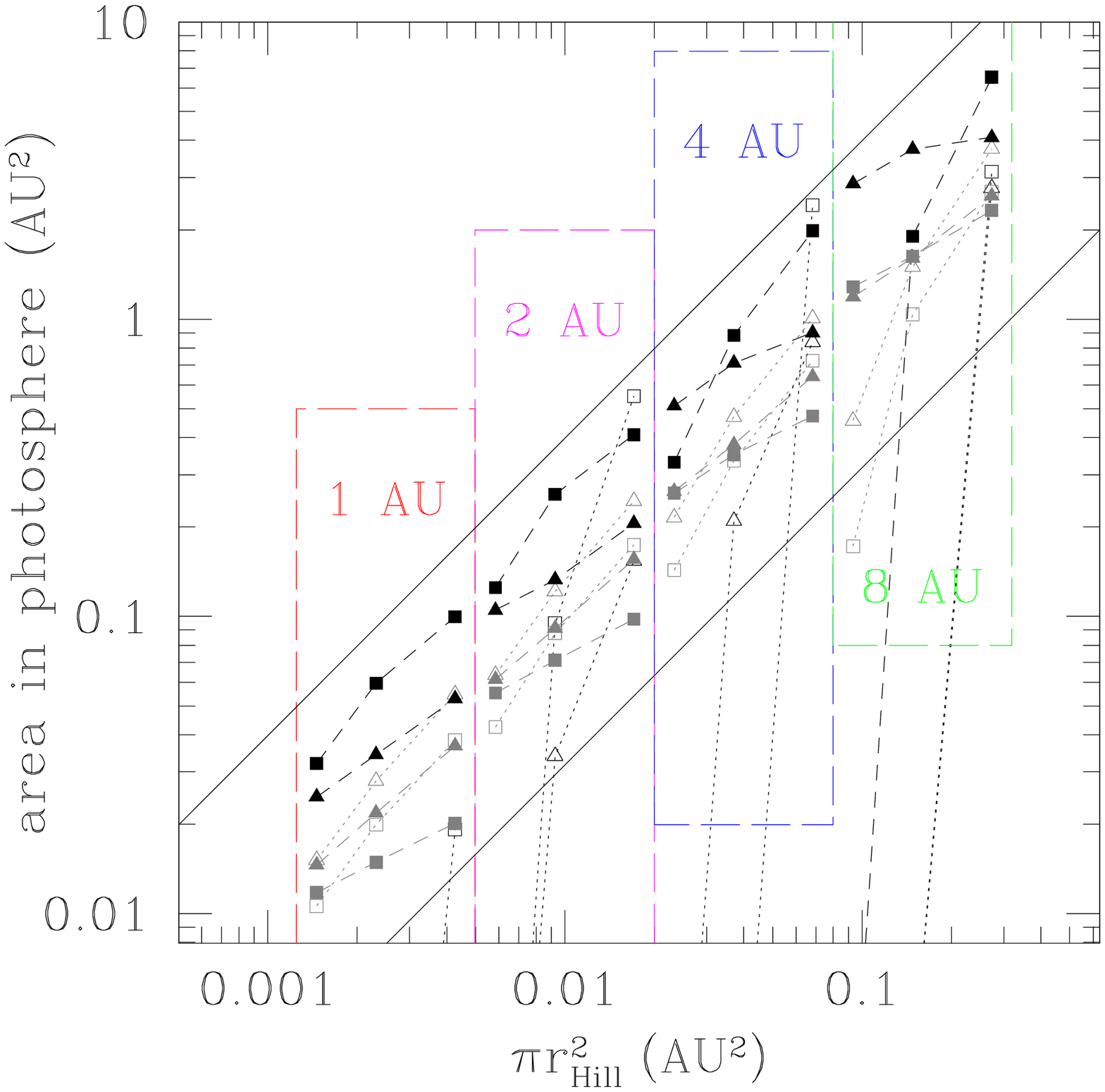}
\caption{\label{hotcoldareas}Areas of heated/cooled regions in 
the photosphere in current model (black) compared to 
results of previous model (grey).  Horizontal axis is the area 
of a circle with radius of $r\sub{hill}$.  Planets with 10, 20 and 
50 M$_{\earth}$ at 1, 2, 4, and 8 AU are shown.  The distances 
for each set of planets are indicated by the long-dashed boxes.   
Heated and cooled regions are represented by 
triangles and squares, respectively.  Filled symbols connected by 
dashed lines show the areas 
heated/cooled beyond half the maximum temperature deviation in the 
photosphere.  Open symbols connected by dotted lines
show the areas that are heated/cooled 
beyond $10\%$ above or below the unperturbed temperature.
The diagonal solid lines show a linear slope.  
}
\end{figure}

We define the area of a spot to be the area of the photosphere 
that is heated or cooled to at least half the maximum temperature 
deviation, excluding the area just above the Hill sphere.  
These spot areas are plotted as filled triangles (hot) or squares
(cold) in \figref{hotcoldareas} connected by dashed lines, 
black for the current models, grey for J-CS.  The temperature 
deviations for the 10 M$_{\earth}$ planet at 8 AU are so 
small that the area of a cold spot is not well-defined.  
The spot masses are generally larger for our results than for the 
JC-S models.  The sizes of cold spots tend to grow faster 
with increasing planet mass than the sizes of hot spots.  

Let us also consider temperature perturbations above or below 
a threshold of 10\%.  
The area of these spots versus the area of a circle with 
radius $r\sub{Hill}$ is plotted as open triangles (hot) or squares
(cold) in \figref{hotcoldareas} connected by dotted lines, 
black for the current models, grey for J-CS.  Lines connected to 
points below the edge of the graph indicate lack of a spot.  
Note that at 1 AU, there are no hot spots above a 10\% temperature 
deviation, consistent with \figref{temp_vs_rhill}.  
The J-CS models produce hot at cold spots for planet parameters.  
At all distances, 10 M$_{\earth}$ planets are too small to 
create either hot or cold spots at the 10\% level in the new models.  
Where cold spots do exist, they are generally larger than 
those produced in J-CS.  However, the hot spots are generally smaller.  

While the temperature deviations predicted from the current models 
are not as great as those predicted in J-CS, they are larger in 
spatial extent.  This may have consequences for observability of 
this phenomenon.  This topic will be addressed in a companion paper.

\section{Conclusion}\label{conclusion}

This paper presents a model for calculating the density and temperature 
perturbations imposed on a protoplanetary disk by an embedded 
protoplanet.  The basic radiative transfer model is adopted from 
Papers I and II, but a number of improvements have been made on that 
work such as density-temperature self-consistency and eliminating 
the assumption of a plane-parallel system.  

We have shown that self-consistently calculating the temperature and 
density significantly increases the effect of planet perturbations 
by means of postive feedback, where shadowed regions cool and contract 
and brightened regions heat and expand.  This demonstrates the 
importance of self-consistency when calculating disk structure 
with radiative transfer.  While it has already been acknowledged 
that stellar irradiation heating is important in setting overall 
disk structure \citep[e.g.][]{CG,thermstab,vertstruct,2004DullemondDominik}, 
this work shows that it is important for considering local perturbations 
on the disk.  

Another important result of this paper is that the temperature 
structure of the disk is extremely sensitive to the angle of incidence 
of stellar irradiation at the surface.  The precise determination 
of the surface of the disk is critical to accurately calculating 
the temperature structure of the disk.  In previous work we had assumed that 
the surface of constant density was a sufficiently good approximatation 
for the surface, but in this work we show that that overestimated the 
temperature perturbation to the disk.  

We have omitted some important physics in this calculation.  We do not 
include heating from accretion onto the planet.  We consider the 
embedded planet to act simply as a gravitational point mass, and 
focus only on the vertical component of gravity.  We do not 
account for non-linearities 
such as spiral density waves.  All these effects are more adequately 
addressed using a three-dimensional hydrodynamic simulation of a planet 
embedded in a disk of gas.  However, since simulations of this sort 
focus on the bulk flow of gas at the midplane, they typically 
have insufficient resolution above the midplane to accurately calculate the 
surface of the disk
\citep[e.g.][]{bate,2004MNRAS.350..829P,2006KlahrKley,2007Oishi_etal}.  
Calculation of radiative transfer, 
even without iterating for self-consistency, is very computationally 
intensive.  To do this iteratively and coupled with three-dimensional 
hydrodynamics is challenging, but is the next logical step to 
improving the accuracy of our results.

\acknowledgements
This work was supported 
by the NASA Astrobiology Institute under
Cooperative Agreement NNA04CC09A.  
Thanks also go to an anonymous referee for helpful suggestions for 
improving this paper.  

\bibliographystyle{apj}
\bibliography{}

\end{document}